\documentclass[5p,twocolumn,10pt,times]{elsarticle}
\usepackage[utf8]{inputenc}
\usepackage{amsmath}
\usepackage{hyperref}
\usepackage{amssymb}
\usepackage{caption}
\usepackage{algorithm}
\usepackage{algorithmicx}
\usepackage{algpseudocode}
\usepackage{amsmath}
\usepackage{color}
\usepackage{multirow}
\usepackage{graphicx}
\usepackage{subfigure}

\newcommand{\ldn}[1]{\textcolor{black}{#1}}
\newcommand{\ldnc}[1]{\textcolor{black}{#1}}

\begin{document}
\baselineskip11pt

\begin{frontmatter}

\title{Deep Feature-preserving Normal Estimation for Point Cloud Filtering}

\author[1]{Dening Lu\fnref{*}}
\author[2]{Xuequan Lu\fnref{*}}
\author[1]{Yangxing Sun}
\author[1]{Jun Wang}

\fntext[*]{Dening Lu and \href{www.xuequanlu.com}{Xuequan Lu} are joint first author. Our programs will be made publicly available. Visit https://github.com/DN-Lu and www.xuequanlu.com for more information.}
\address[1]{Nanjing University of Aeronautics and Astronautics, P.R. China}
\address[2]{Deakin University, Australia}

\date{December 2019}

\begin{abstract}
Point cloud filtering, the main bottleneck of which is removing noise (outliers) while preserving geometric features, is a fundamental problem in 3D field. The two-step schemes involving normal estimation and position update have been shown to produce promising results. Nevertheless, the current normal estimation methods including optimization ones and deep learning ones, often either have limited automation or cannot preserve sharp features. In this paper, we propose a novel feature-preserving normal estimation method for point cloud filtering with preserving geometric features. \ldn{It is a learning method and thus achieves automatic prediction for normals.} For training phase, we first generate patch based samples which are then fed to a classification network to classify feature and non-feature points. We finally train the samples of feature and non-feature points separately, to achieve decent results. Regarding testing, given a noisy point cloud, its normals can be automatically estimated. For further point cloud filtering, we iterate the above normal estimation and a current position update algorithm for a few times. Various experiments demonstrate that our method outperforms state-of-the-art normal estimation methods and point cloud filtering techniques, in terms of both quality and quantity.

\end{abstract}
\end{frontmatter}

\section{Introduction}
\label{sec:introduction}

Point cloud filtering has attracted noticeable attentions recently. It has a wide range of applications, including further geometry processing, computer animation, rendering, computer aided design and so on. Point cloud filtering has remained a challenge to date, \ldn{due to the main bottleneck of noise/outliers removal as well as the preservation of geometric features.}

Existing point cloud filtering techniques can be generally categorized into position based and normal based methods. Position based methods \cite{lipman2007parameterization,huang2009consolidation,preiner2014continuous,yu2018ec,roveri2018pointpronets,rakotosaona2019pointcleannet}, including learning  techniques \cite{yu2018ec,roveri2018pointpronets,rakotosaona2019pointcleannet}, are often not designed to preserve sharp features. Normal based approaches usually contain two common steps: normal estimation and position update. Some recent methods attempted to filter point cloud data by utilizing normal information, such as \cite{sun2015denoising,avron2010,lu2017gpf,lu2018low}. Normals are usually estimated by either the $L_0$/$L_1$/$L_2$/nuclear minimization \cite{sun2015denoising,avron2010,huang2013edge,lu2018low}, or deep learning \cite{guerrero2018pcpnet}. Regarding position update based on estimated normals, the $L_0$/$L_1$/$L_2$ optimization have been often employed \cite{sun2015denoising,avron2010,lu2018low}. However, the $L_0$/$L_1$/nuclear minimization is usually complicated and slow due to their mathematical nature. The $L_2$ optimization such as bilateral normal smoothing \cite{huang2013edge,oztireli2009feature} may sometimes over-sharpen geometric features. Moreover, the optimization methods usually involve multiple parameters, and require tedious tuning to achieve decent results. Learning methods indeed provide alternatives for automatic learning and prediction. Nevertheless, to our knowledge, none of the existing learning methods are introduced to estimate feature-preserving normals which are of particularly great importance to CAD-like models.

\begin{figure*}[t]
  \centering
  \includegraphics[width=\linewidth]{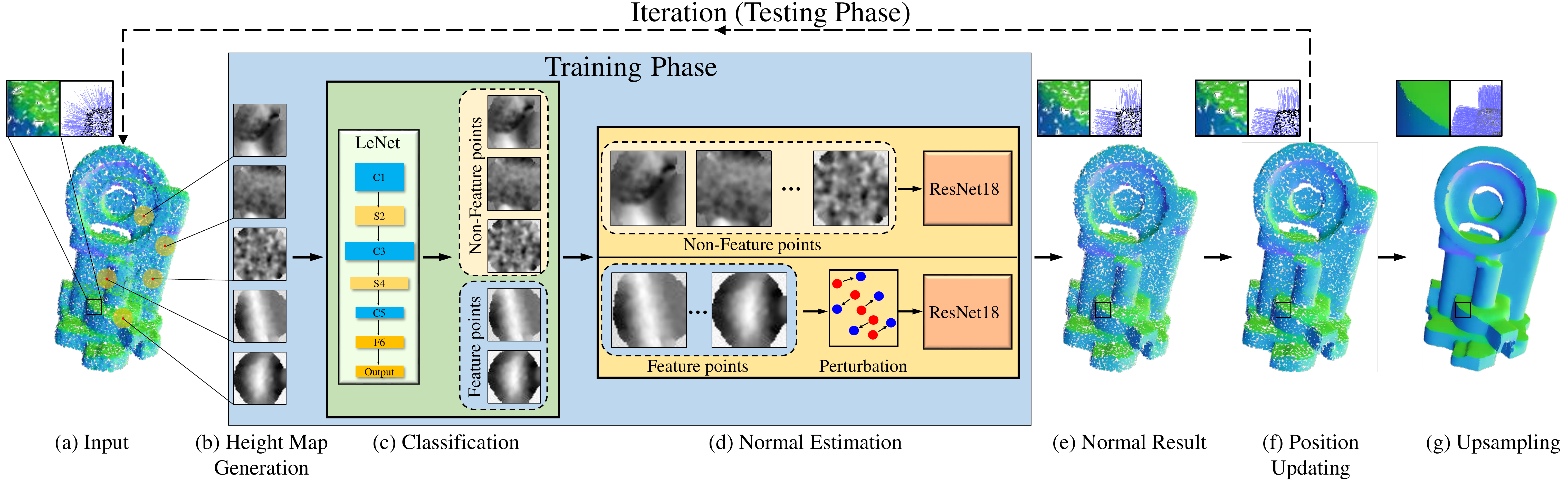}
  \caption{Overview of our approach. The training phase is composed of (a)-(d) for normal estimation. The testing iterates (a)-(f) for a few times to achieve point cloud filtering. (g) is the upsampled result.
  \label{fig:overview}}
\end{figure*}

Motivated by the above issues, in this paper we propose a novel feature-preserving normal estimation method for point cloud filtering. Our core idea is to estimate feature-preserving normals in a deep learning way, and then update point positions with the estimated normals. To achieve this, we propose a framework consisting of three steps for training. We first generate training data, and then classify points into feature points (i.e., anisotropic surface points) and non-feature points (i.e., isotropic surface points) via the classification network. The normal estimation network is then trained on feature points and non-feature points, respectively. In testing phase, we can automatically get normal estimations given an input point cloud. For filtering purpose, the estimated normals are then used to update point positions \cite{lu2018low}. We iterate the normal estimation and position update for a few times to obtain better filtering results. 
To our knowledge, it is the first method to achieve feature-preserving normal estimation via deep learning. Figure \ref{fig:overview} shows an overview for the training and testing phases of the proposed approach.

Various experiments validate our method, and demonstrate that it outperforms the state-of-the-art normal estimation methods and point cloud filtering approaches both visually and quantitatively. The \textit{main contributions} of this paper are:
\begin{itemize}
    \item a novel point cloud filtering framework with preserving geometric features;
    \item a novel method for feature-preserving normal estimation;
    \item a novel method for the classification of feature and non-feature points.
\end{itemize}

\section{Related Work}
\label{sec:relatedwork}

\textbf{Normal Estimation.}
The early research on normal estimation was based on Principal Component Analysis (PCA)~\cite{hoppe1992surface}, which computes a tangent plane for each point according to its neighborhood.
To improve the robustness and accuracy of PCA normals,
several variants have been proposed~\cite{klasing2009comparison,mitra2003estimating,pauly2002efficient,yoon2007surface}.
Other approaches based on Voronoi cells or Voronoi-PCA ~\cite{amenta1999surface,merigot2010voronoi,dey2006provable,alliez2007voronoi}
were also introduced to estimate normals.
To preserve sharp features,
~\cite{sun2015denoising, avron2010} were proposed to minimize the $L_{0}$ or $L_{1}$ norm,
which leads to expensive computation.
Based on the robust statistics and point clustering in a local structure,
~\cite{li2010robust,zhang2013point,liu2015quality} estimated better normals for edges and corners. \ldn{Owing to the ambiguity of the points which locate at the intersection of multi surfaces, recently a pair consistency voting strategy was presented by \cite{zhang2018multi} to compute multiple normals for each feature point. By measuring how close each point is to sharp features, \cite{zhang2018multi} screened the candidate points whose neighbors may be from several different surfaces. Then, a series of reliable tangent planes were fitted to describe the structure of the neighborhood. Conceivably, the normals of these tangent planes were used as the candidate point’s multiple normals. }

\textbf{Point Cloud Filtering.}
\cite{levin1998approximation,alexa2003computing} introduced Moving Least Squares (MLS) methods,
which are designed for smooth surfaces. Such methods may blur sharp features of CAD-like models.
To preserve sharp features,
many improved versions~\cite{adamson2006point,guennebaud2007algebraic,chen2017multi,oztireli2009feature} were proposed.
Based on robust local kernel regression,
robust implicit moving least squares (RIMLS)~\cite{oztireli2009feature} achieves better results than the original MLS.
~\cite{lipman2007parameterization} introduced locally optimal projection without any local parametric representation, and then its variants~\cite{huang2009consolidation,preiner2014continuous} emerged.
Such methods usually contain two energy terms (a data term and a repulsion term) which aim to ensure the geometric fidelity and even distribution of the projected points.
To keep sharp features better, anisotropic LOP~\cite{huang2013edge} was introduced. The $L_1$ minimization was proposed to reconstruct sharp point set surfaces \cite{avron2010}.
Sun et al.~\cite{sun2015denoising} designed a filtering method by applying $L_{0}$ minimization, but it needs a post-processing step to refine the results.
Wu et al.~\cite{wu2015deep} formulated point set consolidation, skeletonization, and completion into a unified framework.
Inspired by GMM, a feature-preserving point set filtering (GPF) method~\cite{lu2017gpf} was proposed, which also contains two terms and takes normal information into consideration to  preserve sharp features.
Non-local methods were also developed from the image filtering field.
\cite{digne2012similarity} proposed a denoising method based on local similarities,
which smooths a height vector field by comparing the neighborhood of a point with neighborhoods of other points on the surface.
Oussama et al.~\cite{remil2017data} built a prior library based on the similar patches,
and formulated a global, sparse optimization problem to enforce selecting representative priors.
\ldn{Building upon the Poisson integration model, ~\cite{bahr2017fast} used an iterative Krylov subspace solver to achieve surface normal integration for 3D reconstruction, and performed a thorough numerical study to identify an appropriate preconditioner. This method generates satisfactory reconstructions with low computational time and low memory requirements. \cite{li2011globfit} recovered a set of locally fitted primitives along with their global mutual relations for man-made objects. The method achieved desired reconstructions for CAD-like models, since it considers both local primitive fitting and global relations.} Lu et al.~\cite{lu2018low} proposed a low rank matrix approximation approach for filtering point clouds and meshes.
Chen et al.~\cite{chen2019multi} presented a multi-patch collaborative method which transformed irregular local point cloud patches to regular local height-map patches.

\textbf{Learning-based Methods.}
There exist a few deep learning based methods for normal estimation and point cloud filtering.
In terms of normal estimation, the early work based on a 2D CNN was introduced~\cite{boulch2016deep}, which transforms a 3D patch into a 2D Hough space accumulator and formulates it as a regression problem.
Based on the PointNet~\cite{qi2017pointnet}, PCPNet~\cite{guerrero2018pcpnet} has been designed as a deep multi-scale architecture, which leads to a better normal estimation performance.
Later, Ben-Shabat et al.~\cite{ben2019nesti} presented a mixture- of-experts (MOE) architecture, Nesti-Net, which relies on a data driven approach for selecting the optimal scale around each point and encourages sub-network specialization.

\ldn{In terms of point cloud filtering, EC-Net~\cite{yu2018ec}, PointProNet~\cite{roveri2018pointpronets}, PointcleanNet \cite{rakotosaona2019pointcleannet} and Pointfilter \cite{zhang2020pointfilter} were proposed recently. In detail, EC-Net designed a joint loss function to realize the edge-aware effects. It is suitable for dense point clouds whose scale of structure is invariant. Moreover, EC-Net~\cite{yu2018ec} needed tedious manual efforts in sharp edge labeling. PointProNet \cite{roveri2018pointpronets} used a heightmap generation network to convert the unordered points to regularly height maps, and then employed the CNN architecture to consolidate the point cloud. PointcleanNet \cite{rakotosaona2019pointcleannet} proposed a two-stage data-driven denoising approach, which discarded outlier samples firstly and then estimated the local properties for the remaining points. Pointfilter \cite{zhang2020pointfilter} is position based and depends on prior normal information for training. However, nearly none of these methods are designed to estimate feature-preserving normals and further point cloud filtering.}
\section{Overview}
\label{sec:overview}

Figure~\ref{fig:overview} shows an overview of our proposed method, which consists of the training and testing phases. Training includes data generation, classification and normal estimation. In testing phase, apart from the three steps above, an additional step (i.e., position update) is included to match the estimated normals and achieve the goal of point cloud filtering.

\textbf{Data generation.}
Taking as input a noisy point cloud, we represent each point and its neighbors as a 2D height map, to meet the classical CNN requirement. Specifically, we present a simple projection approach based on PCA. 
Since our method consists of two networks (i.e., classification and normal estimation), we define two different labels for each point: feature/non-feature label and normal label.

\textbf{Classification.}
To improve the accuracy of the normal estimation network, we first distinguish the feature and non-feature points, which is formulated as a regression problem. 
Instead of the commonly used $l_{2}$ loss, we introduce a weighted $l_{2}$ loss, leading to a better classification result for geometric features.

\textbf{Normal estimation.}
Feature points are assumed to own multiple normals with weights, inspired by \cite{zhang2018multi}.
We train the normal estimation network for feature points and non-feature points, respectively, to achieve better normal estimations than the mixed training.

\textbf{Position update.}
At test phase, we first predict the normals of the input point cloud, and then employ a position update algorithm to update point positions \cite{lu2018low}. To achieve better results, the normal prediction and position update are alternately called for a few iterations.

\section{Data Generation}
\label{sec:step1}

Given a noisy point cloud $\mathbb{P}=\left \{ p_{i} \right \}_{i=1}^{N}\subset R^3$, we define each point $p$ ($3\times1$ vector) with its local neighborhood (spherical search) as a local patch $\chi$. Since the classical CNN requires 2D input, $\chi$ needs to be transformed to a 2D representation. Height map is an intuitive choice which has been proven to be effective for deep learning networks \cite{roveri2018pointpronets}, and thus we use it in this work.

\begin{figure}[hbp]
  \centering
  \includegraphics[width=\linewidth]{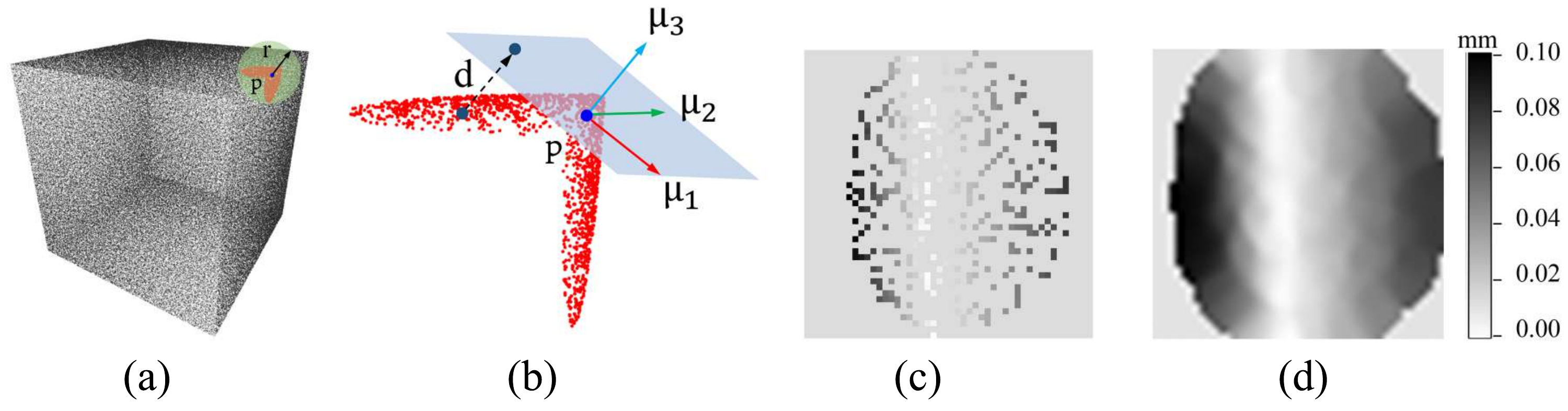}
  \caption{Height map generation based on PCA. (a) The local patch for the point $p$. (b) The projection procedure. (c) The original height map. (d) The final height map after the Gaussian-weight interpolation.
  \label{fig:heightmap_generation}}
\end{figure}

\textbf{Height map generation.}
As shown in Figure \ref{fig:heightmap_generation}, we present a simple yet effective projection scheme based on PCA, for the generation of height maps. Denote the eigenvalues of the position covariance matrix $C_{p}$ (Eq. \eqref{eq:pca}) for point $p$ by $\lambda _{1}$, $\lambda _{2}$ and $\lambda _{3}$ ($\lambda _{1}\geq \lambda _{2}\geq \lambda _{3}$), with the corresponding eigenvectors $\mu _{1}$, $\mu _{2}$ and $\mu _{3}$. 
\begin{equation}
\label{eq:pca}
C_{p}=\frac{1}{|\chi|}\sum_{p_{i}\in \chi}(p-p_{i})(p-p_{i})^T,
\end{equation}
where $|\chi|$ is the number of $p$'s neighboring points.
We take the plane $P_{(p,\mu _{3})}$ as the projection plane with a 2D regular
grid (size $M=m\times m=48\times48$ cells), where the $x$-axis and $y$-axis are $\mu _{1}$ and $\mu _{2}$, respectively. Given a point $p_{i}$ in $\chi$, we denote its corresponding cell in the the projection plane as $c_{i}$, and the coordinates $(x,y)_{i}$ and value $H(c_i)$ of $c_{i}$ are defined as:
\begin{equation}
c_{i}=(x,y)_{i}=\left ( \frac{( p_{i}-p )\cdot \mu _{1}+r}{2r} m,\frac{( p_{i}-p )\cdot \mu _{2}+r}{2r} m \right ),
\end{equation}
\begin{equation}
H(c_i)= (p_{i}-p) \cdot \mu _{3},
\end{equation}
\ldn{where $r$ is the radius of the patch $\chi$, and is empirically set to $5$ times the average distance $r_{average}$ of the input point cloud $\mathbb{P}$. This enables an adaptive radius based on the average distance for each input point cloud, as suggested by \cite{roveri2018pointpronets}. } If a cell has more than one points, we simply choose the one with the shortest projection distance.

Since the number of the points in $\chi$ is generally less than $M$, we apply the Gaussian-weight interpolation to fill up the vacant cells (i.e., no points). Other types of interpolation may be possible and we found the Gaussian-weight interpolation works very well.
\begin{equation}
\label{eq:interpolation}
H_{g}(c_i) =
\left\{
             \begin{array}{lr}
             \frac{1}{w_i}\sum_{c_k \in \mathbb{N}_{c_i}} g(c_i,c_k)H(c_k),   \text{if~}  |\mathbb{N}_{i}|\neq 0 &  \\
             0,  \text{otherwise}&
             \end{array}
\right.
\end{equation}
where $w_i= \sum_{c_k \in \mathbb{N}_{c_i}} g(c_i,c_k)$, and $g(c_i,c_k)=e^{-\frac{|| c_{i}-c_{k} ||^{2}}{\sigma_{g} ^{2}}}$. $\mathbb{N}_{c_i}=\left \{ c_{k}|~|| c_{i}-c_{k} ||<\eta  \right \}$, where $\eta$ means the radius of the Gaussian-weight interpolation. $\eta$ and $\sigma_{g} $ are set to $\frac{m}{6}$ and $\frac{\eta}{2.5}$, respectively.

\begin{figure}[htbp]
  \centering
  \includegraphics[width=\linewidth]{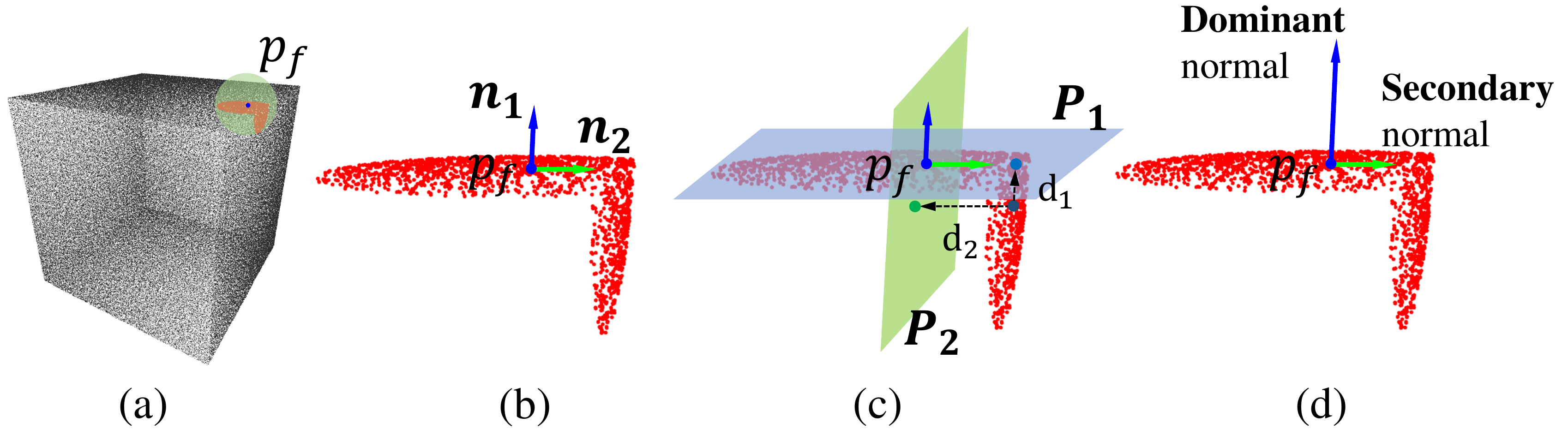}
  \caption{\ldn{Multiple normals for a feature point. (a) The local patch for the feature point $p_{f}$. (b) The multiple normals of $p_{f}$, where different color arrows represent different normals. (c) The priority calculation for the multiple normals, where the plane $P_1$ is defined by $p_f$ and the blue normal $n_1$, and the plane $P_2$ is defined by $p_f$ and the green normal $n_2$. The smaller distance $d_{1}$ would lead to a greater contribution. (d) The dominant and secondary normal of $p_{f}$ based on the priority calculation (Eq. \eqref{eq:weight1} and Eq.\eqref{eq:weight2}). }
  \label{fig:weight_normal}}
\end{figure}

\begin{figure}[htbp]
  \centering
  \includegraphics[width=\linewidth]{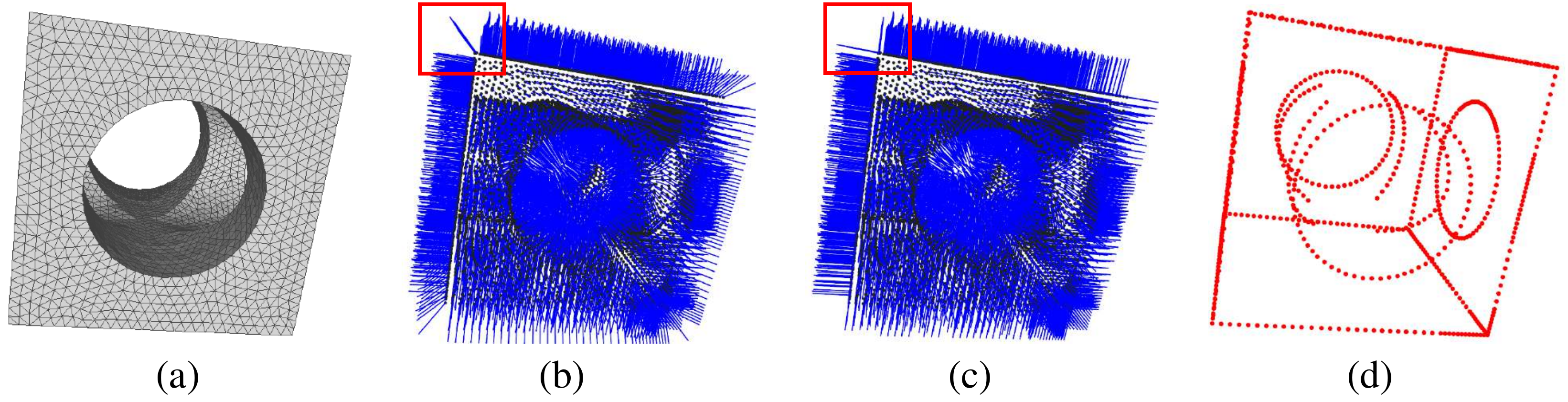}
  \caption{\ldn{Ground-truth feature points extraction. (a) The triangle mesh model. (b) normals by \cite{thurrner1998computing}. (c) Feature-preserving normals by the latter scheme. (d) The extraction of ground-truth feature points.
  }
  \label{fig:gt_feature}}
\end{figure}

\begin{figure}[htbp]
  \centering
  \includegraphics[width=\linewidth]{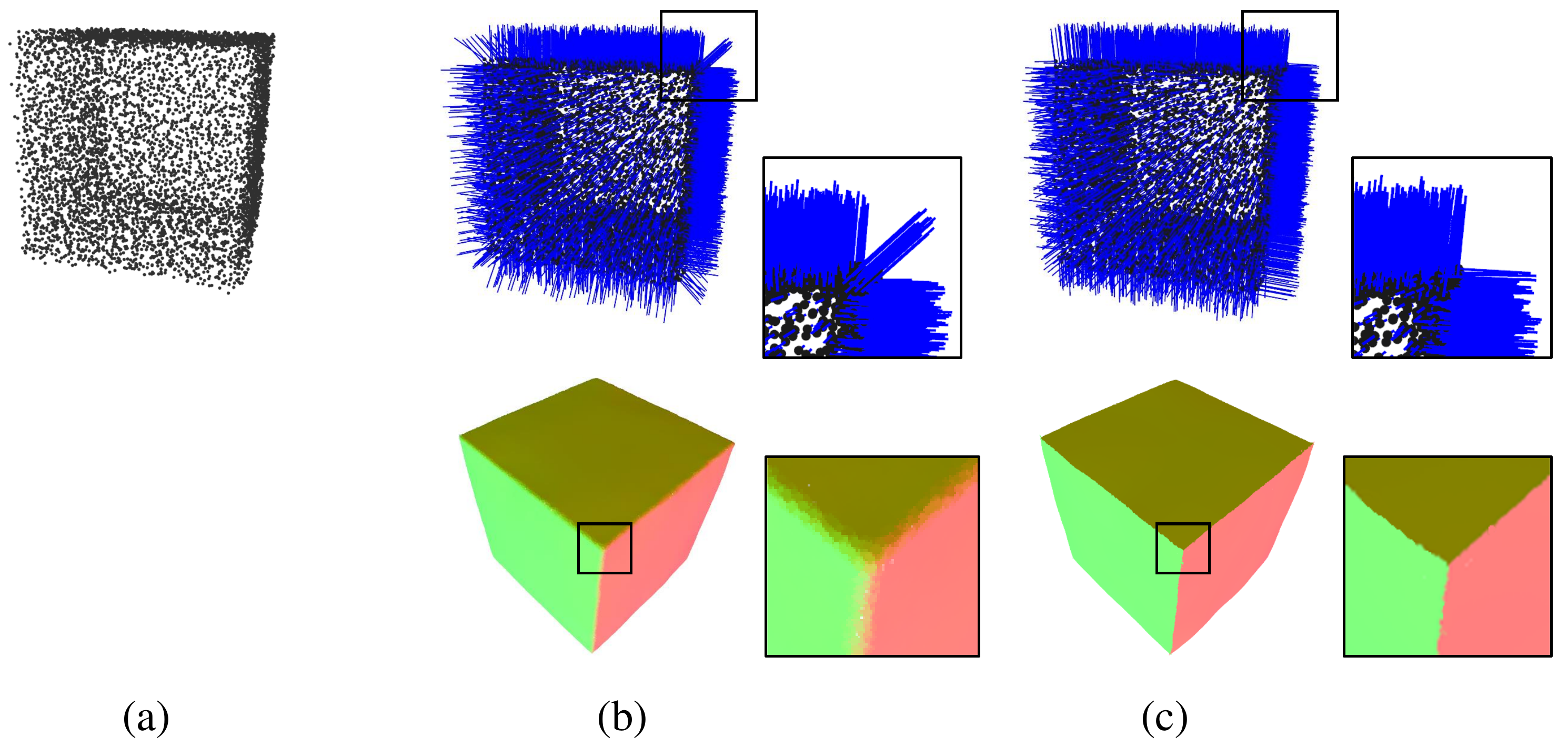}
  \caption{\ldn{(a) Input (1.5$\%$ noise). (b) normals by \cite{thurrner1998computing} and filtering result. (c) feature-preserving normals by the latter method and filtering result.
  }
  \label{fig:45_normal}}
\end{figure}

\textbf{Training labels.}
We need to define the ``label'' for each point, for training purposes. To improve the accuracy of the normal estimation network, we first classify all points in $\mathbb{P}$ into feature points (anisotropic surface points) and non-feature points (isotropic surface points), and feed them into the normal estimation network for separate training. \ldn{For labeling feature points and non-feature points, we define the feature point set $\Re$ and non-feature point set $\Im$ from $\mathbb{P}$ as:}
\begin{equation}
\label{eq:label_gt1}
\Re =\left \{ p_{i} \in \mathbb{P} \mid  || p_{i}-p_{g}||<r_{f}, \forall p_{g}\in \Psi \right \}, 
\end{equation}
\begin{equation}
\label{eq:label_gt2}
\Im =\left \{ p_{i} \in \mathbb{P} \mid  || p_{i}-p_{g}|| \geq r_{f}, \forall p_{g} \in \Psi \right \}, 
\end{equation}
where $\Psi$ is the ground-truth feature points set corresponding to $\mathbb{P}$, $p_{g}$ is a ground-truth feature point, and $r_{f}$ is the distance threshold which defaults to $2$ times the average distance $r_{average}$.
\ldn{The ground-truth feature points ($\Psi$) are located easily and automatically. We generate the ground-truth point clouds and corresponding normals from the triangle meshes (Sec. \ref{sec:results}). We employ two approaches to compute the normals of vertices (i.e., points), to identify feature points. One is based on Thurmer et al. \cite{thurrner1998computing}, which computes the normal as a weighted sum of the incident face normals, and the other takes one of the incident face normals as the normal of the vertex directly, as shown in Figure~\ref{fig:gt_feature}. For the non-feature points (i.e., isotropic area), the two computed normals are very similar.
However, the resulting normals are noticeably different for the feature points, such as edge or corner points, which is illustrated in Figure~\ref{fig:gt_feature}(b) and \ref{fig:gt_feature}(c). As with \cite{lu2018low}, we regard the normals by the latter method as  \textit{feature-preserving normals} (Figure \ref{fig:gt_feature}(c)), which is also to facilitate the position update (Sec. \ref{sec:step4}) in terms of point cloud filtering (Figure~\ref{fig:45_normal}(c)). By contrast, the former method would blur the sharp features, as shown in Figure~\ref{fig:45_normal}(b). As a result, we can easily detect and extract the ground-truth feature points by setting a threshold ($18^{\circ}$ by default) for the angle between the two normals calculated by the two above schemes. After this, we can simply extract the feature and non-feature points for classification training via Eq. \eqref{eq:label_gt1} and Eq. \eqref{eq:label_gt2}.}

Since our framework involves two networks (classification network and normal estimation network), we need to define labels for each of them.
Regarding the classification network, we take vectors $\tilde{\gamma_{1}}=(1,0)$ and $\tilde{\gamma_{2}}=(0,1)$, to represent the labels of an extracted feature point and an extracted non-feature point, respectively. Note that the vector as the label accounts for the incidences of a point for being both a feature point and a non-feature point simultaneously, and works well in this work. Regarding the normal estimation network, we introduce the normal label for the non-feature points and feature points, respectively. For a non-feature point, we take the normal of its nearest point in ground truth as the normal label. The normals of feature points, such as edge points and corner points, are undefined or ambiguous, while the multiple normals \cite{zhang2018multi} have been proposed for a clear definition. Inspired by the multi-normal concept \cite{zhang2018multi}, we define the multiple normals (default to two normals) for each feature point, as demonstrated in Figure~\ref{fig:weight_normal}. 
Given a feature point $p_{f}$ and its defined local neighbors in $\mathbb{P}$, we select two corresponding ground-truth normals with the largest angle, as the multiple normals of $p_{f}$ (denoted as $n_{1}$ and $n_{2}$). These two normals also act as the normal label ($6\times 1$ vector) for this feature point.

In testing phase, we need to get a normal ($3\times 1$ vector) for a feature point, rather than a $6\times 1$ vector. To do so, we first define the priorities of the two normals for training.
Specifically, for the feature point $p_{f}$ and its neighborhood $\chi_{f}$, we define two planes $P_{1}=P_{(p_{f},n_{1})}$ and $P_{2}=P_{(p_{f},n_{2})}$ based on the multiple normals $n_{1}$ and $n_{2}$. Based on this, we can calculate the priorities $\omega_1$ and $\omega_2$ of $n_{1}$ and $n_{2}$ as:
\begin{equation}
\label{eq:weight1}
\omega_{1}=\sum_{p_{i} \in \chi_{f}}\frac{e^{\frac{-d(p_{i},P_{1})^{2}}{\sigma_{f} ^{2}}}}{e^{\frac{- d(p_{i},P_{1})^{2}}{\sigma_{f} ^{2}}}+e^{\frac{-d(p_{i},P_{2})^{2}}{\sigma_{f} ^{2}}}},
\end{equation}
\begin{equation}
\label{eq:weight2}
\omega_{2}=\sum_{p_{i} \in \chi_{f}}\frac{e^{\frac{-d(p_{i},P_{2})^{2}}{\sigma_{f} ^{2}}}}{e^{\frac{- d(p_{i},P_{1})^{2}}{\sigma_{f} ^{2}}}+e^{\frac{-d(p_{i},P_{2})^{2}}{\sigma_{f} ^{2}}}},
\end{equation}
where $d(p,P)$ represents the distance between point $p$ and plane $P$, and $\sigma_{f}$ is a scaling parameter, which is empirically set to $2$ times $r_{average}$.

\begin{figure}[h]
  \centering
  \includegraphics[width=\linewidth]{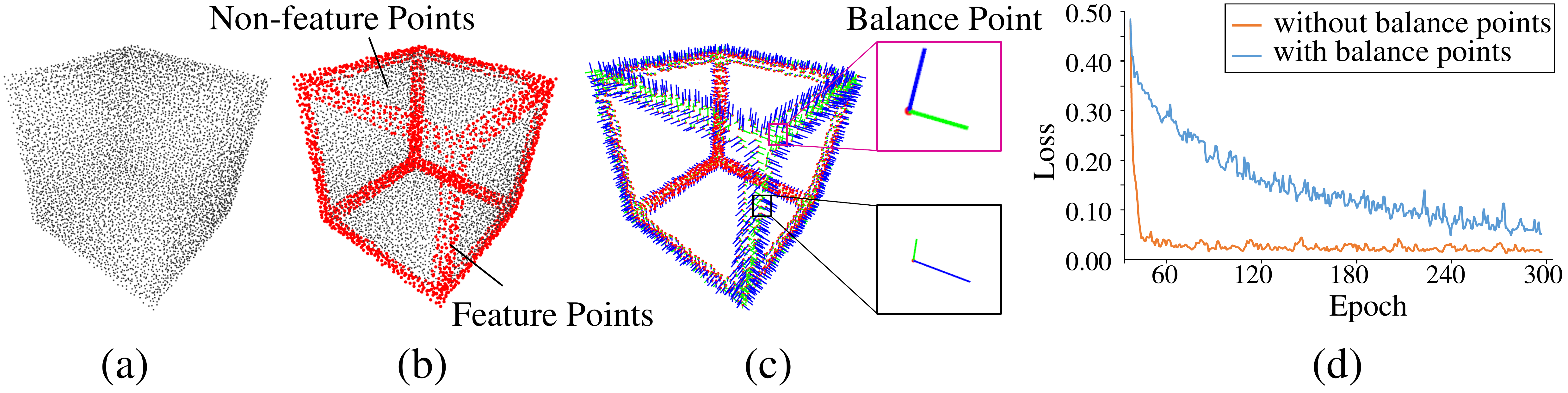}
  \caption{The balance points in the feature point set, whose multiple normals have similar weights. (d) shows the comparison of normal estimation network training with and without balance points.
  \label{fig:multi_normal}}
\end{figure}

\begin{figure}[b]
  \centering
  \includegraphics[width=0.9\linewidth]{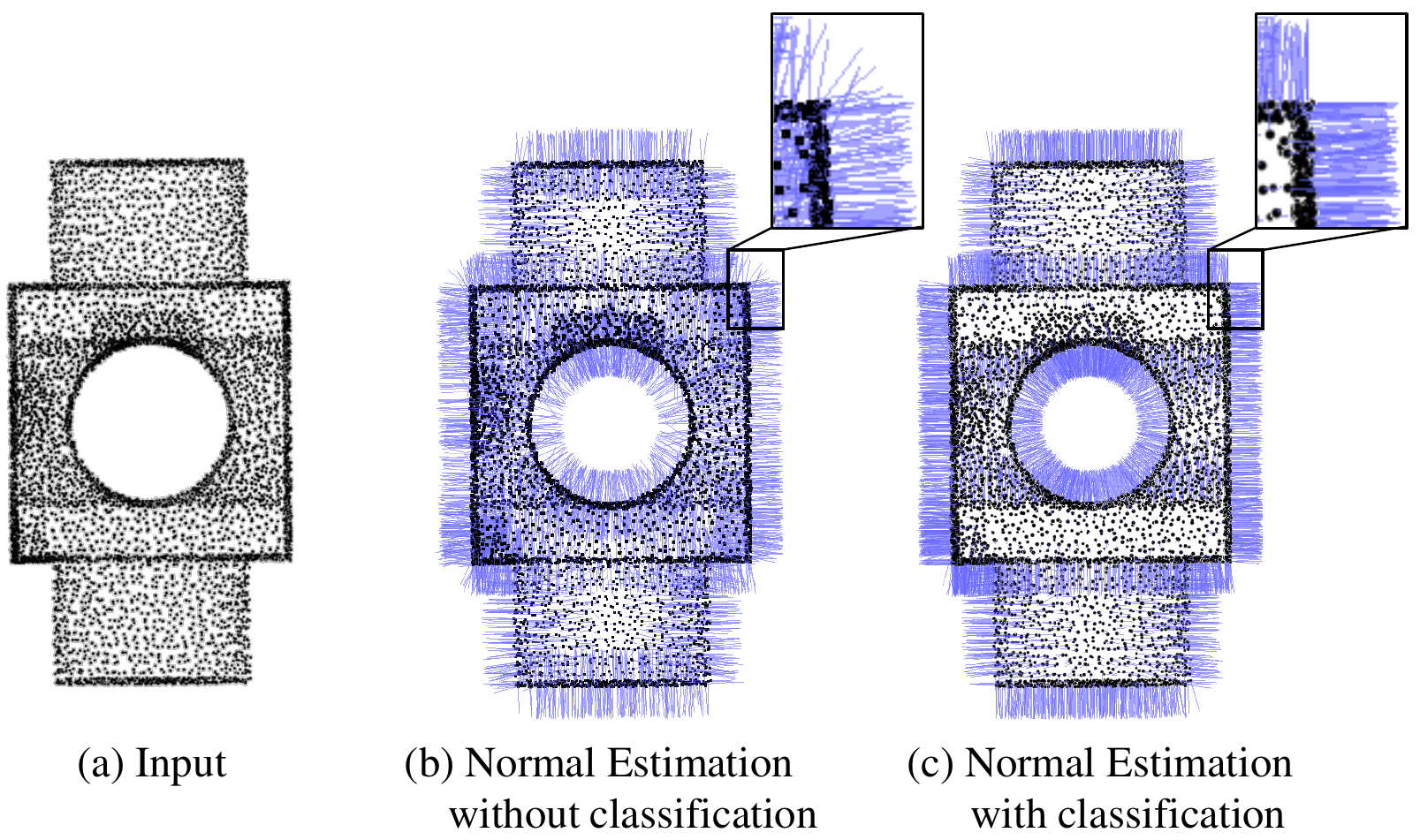}
  \caption{Normal estimation with or without the classification step.
  \label{fig:classification_import}}
\end{figure}

As described in Eq. \eqref{eq:weight1} and \eqref{eq:weight2}, the priorities of the multiple normals are based on the contributions of all points in $\chi_{f}$, and that a smaller distance would lead to a greater contribution. Therefore, the normal with a larger priority is defined as the dominant normal, while the other is chosen as the secondary normal. In testing, the dominant normal is simply selected as the final prediction.
For corners, we still randomly choose two normals with the largest priorities as the multiple normals, even though the corner point may have more than a pair of normals with the same largest angle.
\ldn{Notice that there are some balance points in the feature point set, whose multiple normals have similar weights, as shown in Figure~\ref{fig:multi_normal}. Since the normal priority of a balance point is indistinctive, it may hinder the training of the normal estimation network, as illustrated in Figure~\ref{fig:multi_normal}(d). 
To solve this issue, we simply remove such points from the feature point set.}

\section{Classification}
\label{sec:step2}

We notice that the proportion of feature points is often small, especially for CAD-like models. As a result, the direct training of the normal estimation network based on all points would result in limited accuracy for feature points, as shown in Figure \ref{fig:classification_import}. Therefore, to improve the accuracy, we first classify the feature and non-feature points by training the classification network (this section), and then respectively train the normal estimation network for feature points and non-feature points (Sec. \ref{sec:step3}).

LeNet \cite{lecun1998gradient}, which has been proven to be adaptable to various regression problems in image processing, is chosen for this classification task.
It consists of three convolutional layers, two poolings and two fully connected layers.
In addition, we resize our height maps ($48 \times 48$) to the input size ($32 \times 32$) of LeNet \cite{lecun1998gradient}.
We introduce the following weighted $l_{2}$ loss to train this network: 
\begin{equation}
L_{c}=\sum_{p_{i}\in \Gamma} g(\theta_{i})\cdot||\gamma_{i} -\tilde{\gamma_{i}}||^{2},
\end{equation}
where $\Gamma$ includes all points in the training set for classification, $\gamma_{i}$ and $\tilde{\gamma_{i}}$ are the output of the classification network and the classification label for point $p_{i}$, respectively. $g(\theta_{i})=e^{1-(\frac{\cos{\theta_{i}}}{\cos{\sigma_{\theta}}})^{2}}$ and produces greater weights to feature points for better recognition. $\sigma_{\theta}$ is empirically set to $30^{\circ}$, and $\theta_i$ is defined as:
\begin{equation}
\theta_{i} =
\left\{
             \begin{array}{lr}
             \arccos{(\frac{n_{i1} \cdot n_{i2}}{||n_{i1}||||n_{i2}||})},   \text{if }  p_{i} \in \Re &  \\
             0,  \text{if } p_{i} \in \Im \\
             \end{array}
\right.
\end{equation}
where $n_{i1}$ and $n_{i2}$ are the multiple normals of $p_{i}$. $L_{c}$ is differentiable, so the gradients can be back-propagated through the network. In testing time, by setting a threshold $\varpi_{t}$ (default to $0.85$), the feature and non-feature point set $\mathbb{R}_{f}$ and $\mathbb{R}_{non-f}$ of the test model $\mathbb{R}$ can be naturally obtained:

\begin{equation}
\mathbb{R}_{f}=\left \{ p_{i} \in \mathbb{R} | \varpi_{i} > \varpi_{t} \right \},
\end{equation}
\begin{equation}
\mathbb{R}_{non-f}=\left \{ p_{i} \in \mathbb{R} | \varpi_{i} \leq  \varpi_{t} \right \},
\end{equation}
where $\varpi_{i}$ is the classification score, defined as a proportion style:
\begin{equation}
\varpi_{i} =\frac{a_{i}}{a_{i}+b_{i}},
\end{equation}
where $(a_{i},b_{i})$ is the output of the classification network for $p_{i}$. $\varpi_{i}$ accounts for the ratio between $a_{i}$ and the total $(a_{i}+b_{i})$,  and a greater $\varpi_{i}$ means a greater probability of being a feature point.

\begin{figure}[h]
  \centering
  \includegraphics[width=0.8\linewidth]{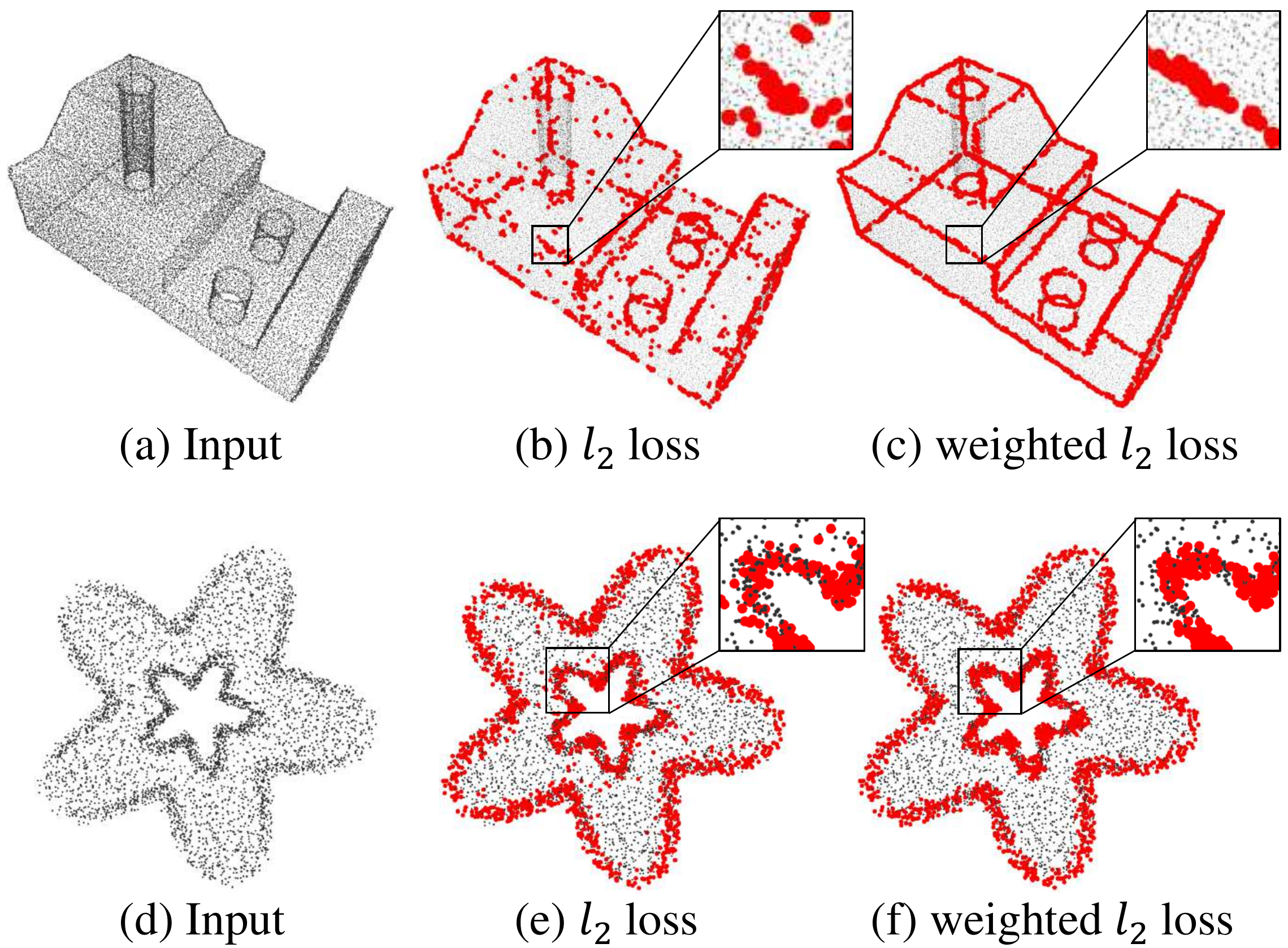}
  \caption{\ldn{Classification results with two different losses, the commonly used $l_2$ loss and our weighted $l_2$ loss. Feature points are colored in red. Classification accuracy ($\%$): (b) $88.5$, (c) $96.6$, (d) $84.8$ and (e) $92.7$.}
  \label{fig:classification_loss_weight}}
\end{figure}

We choose a weighted $l_2$ loss, rather than the commonly used $l_2$ loss, because the latter is inferior to the former in recognizing feature points and non-feature points. Figure \ref{fig:classification_loss_weight} shows the comparisons of the two losses.

\section{Normal Estimation}
\label{sec:step3}

\textbf{Network architecture.}
Our normal estimation network takes ResNet18 \cite{he2016deep} as the backbone. It takes a height map as input, and outputs a $6\times 1$ vector  or $3\times 1$ vector for a feature point and a non-feature point, respectively. For training, the $l_{2}$ penalization is adopted as the loss:
\begin{equation}
L_{n}=\sum_{p_{j}\in \mathbb{Q}} || n_{p_{j}}- \breve{n_{p_{j}}}||^{2},
\end{equation}
where $\mathbb{Q}$ contains feature points or non-feature points in the training set for normal estimation. $n_{p_{j}}$ and $\breve{n_{p_{j}}}$ are the estimated normal(s) and the normal label for point $p_{j}$, respectively. It should be noted that the normal label needs to be transformed to the eigen space of the corresponding local patch in the noisy point cloud before training. It is achieved by multiplying the normal vector by the eigen matrix (formed by the three eigen vectors of Eq. \eqref{eq:pca}). We do this since we found the original normal label would lead to strong oscillation and even non-convergence during training. For testing, the estimated normal can be easily computed with a corresponding inverse matrix to the eigen matrix.
\begin{figure}[t]
  \centering
  \includegraphics[width=\linewidth]{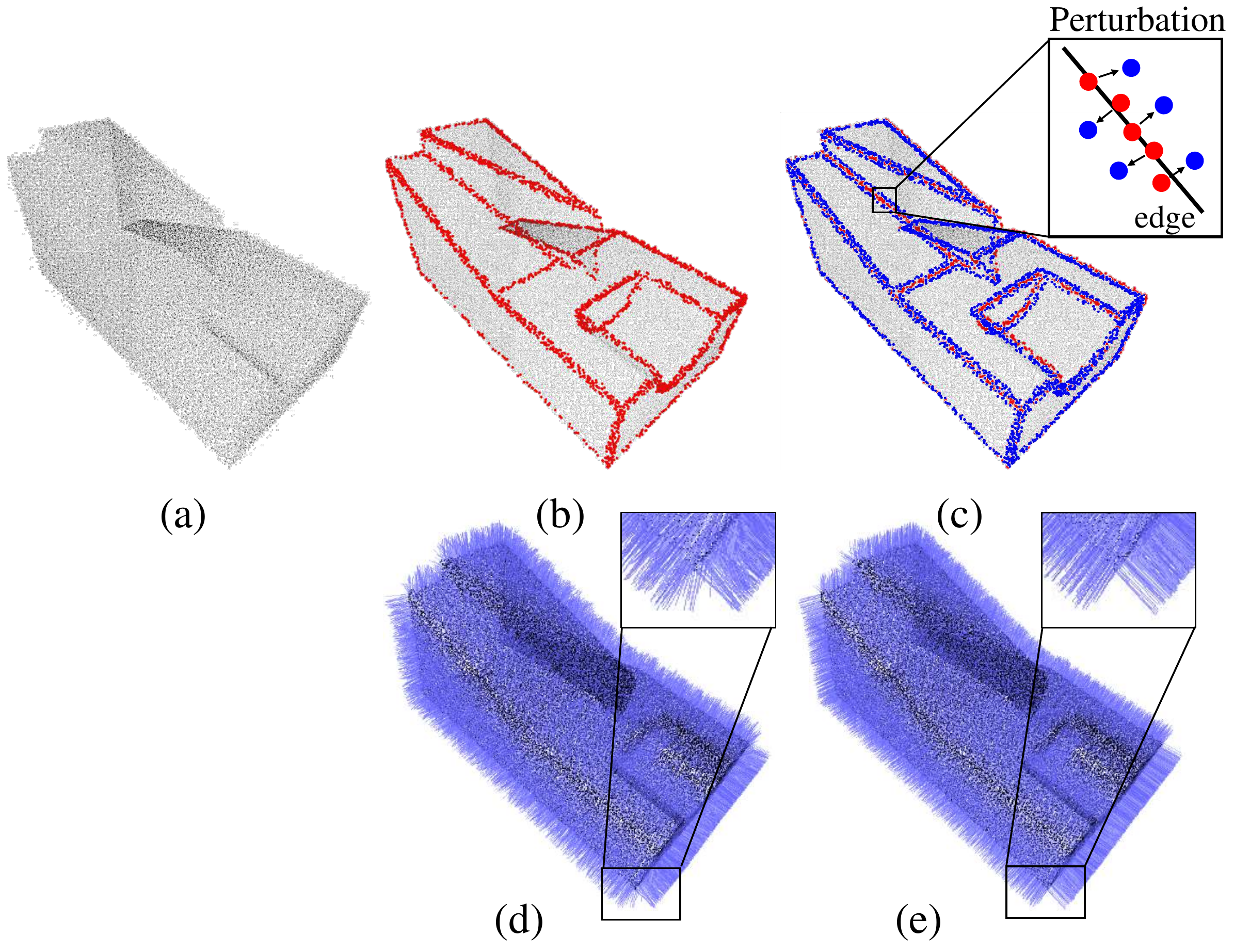}
  \caption{\ldn{Perturbation strategy. (a) Input point cloud. (b) The classification result (feature points in red). (c) The feature points after perturbation (in blue). Classification accuracy ($\%$): (b) $97.8$,
  and the root mean square errors of normal estimation are ($\times 10^{-3}$): (d) $8.2$ and (e) $4.7$.}
  \label{fig:feature_move}}
\end{figure}

\textbf{Perturbation strategy.}
The perturbation strategy is designed for the testing phase only.
After classification, we can obtain the feature point set of the testing point cloud. However, the height maps of the feature points are not expected to be directly fed into the normal estimation network, due to the balance points (Sec. \ref{sec:step1}).
Because it is impossible to calculate the priorities (Eq. \eqref{eq:weight1} and \eqref{eq:weight2}) to judge the balance points, we simply introduce a tiny perturbation to each feature point $p_{f}$, to push it a bit away from the equilibrium status, as demonstrated in Figure~\ref{fig:feature_move}. The feature point after perturbation ($\tilde{p_{f}}$) can be defined as:
\begin{equation}
\tilde{p_{f}}=p_{f} + \varepsilon \cdot \frac{\nu}{|| \nu ||},
\end{equation}
where $\nu$ represents the perturbation direction for $p_{f}$, and is defined as the orientation pointing to the neighboring point with the smallest classification score $\varpi$. $\varepsilon = \varpi \cdot r_{average}$.
Based on $\tilde{p_{f}}$, we re-generate the height map, and feed it to the normal estimation network. \ldnc{\textit{Note that $\tilde{p_{f}}$ is only used to re-generate a height map and $p_{f}$ is thus not changed.}} \ldn{To further demonstrate the performance of the perturbation strategy, we estimate normals with (Figure~\ref{fig:feature_move}(b)) and without perturbation (Figure~\ref{fig:feature_move}(c)). Figure~\ref{fig:feature_move}(d) and \ref{fig:feature_move}(e) show the estimation results, respectively. It is obvious that the latter achieves a better result than the former both visually and quantitatively, since there are balance points in edge and corner parts in Figure~\ref{fig:feature_move}(b). }

\ldn{We do not perform the perturbation strategy for the training phase, since we found the training data already involves rich similar samples to the samples perturbed on balance points. As a result, the decent normal estimations can be achieved.}

\section{Position Update}
\label{sec:step4}

In testing phase, we adopt an efficient point update algorithm ~\cite{lu2018low} to match the estimated normals output by Sec. \ref{sec:step3}.
Assuming that the point $p_{i}$ and its neighboring information $\mathbb{N}_{i}$ are known and unchanged \cite{lu2018low}, its new position can be calculated as:
\begin{equation}
\breve{p_{i}}=p_{i}+\alpha_{i}\sum_{p_{k} \in \mathbb{N}_{i}}\left ( p_{k}-p_{i} \right ) \left ( n_{k}^{T}n_{k}+n_{i}^{T}n_{i} \right ),
\end{equation}
where $\alpha_{i}$ is the step size that is default to $\frac{1}{3 |\mathbb{N}_{i}|}$, as suggested by the authors \cite{lu2018low}.

\begin{figure}[h]
  \centering
  \includegraphics[width=0.9\linewidth]{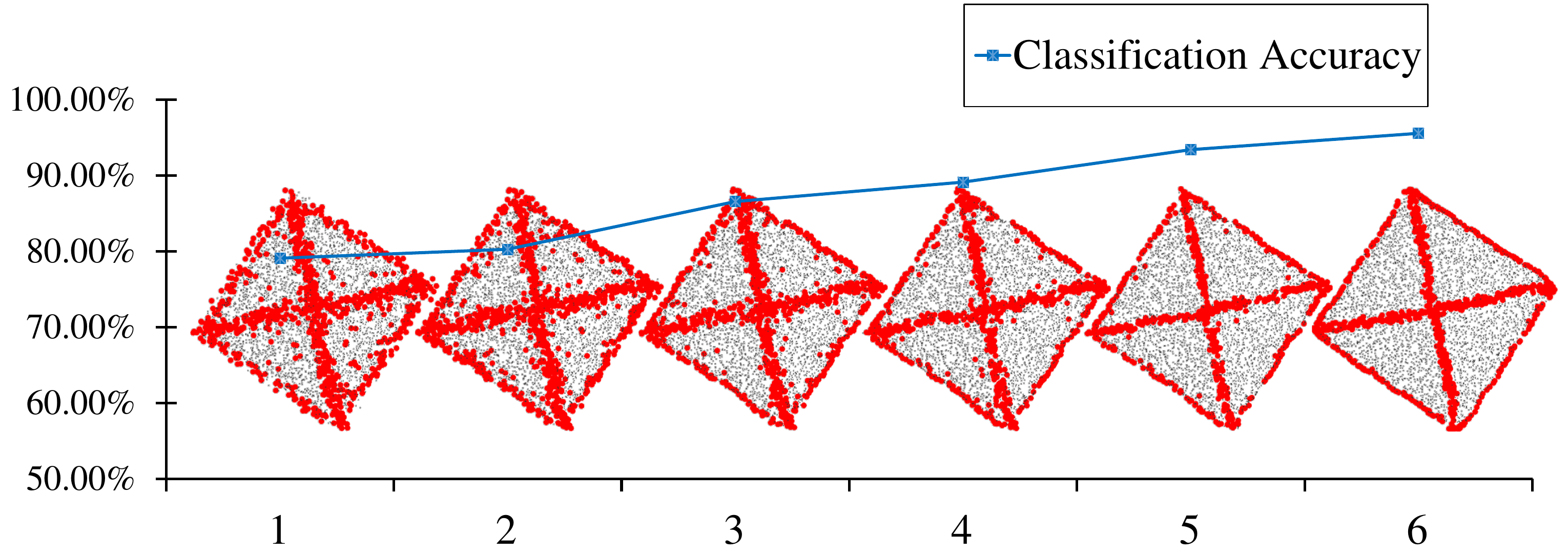}
  \caption{\ldn{The classification results of the Octahedron point set (1.5$\%$ noise) in different iterations.}
  \label{fig:classification_iteration}}
\end{figure}
\begin{figure}[h]
  \centering
  \includegraphics[width=0.9\linewidth]{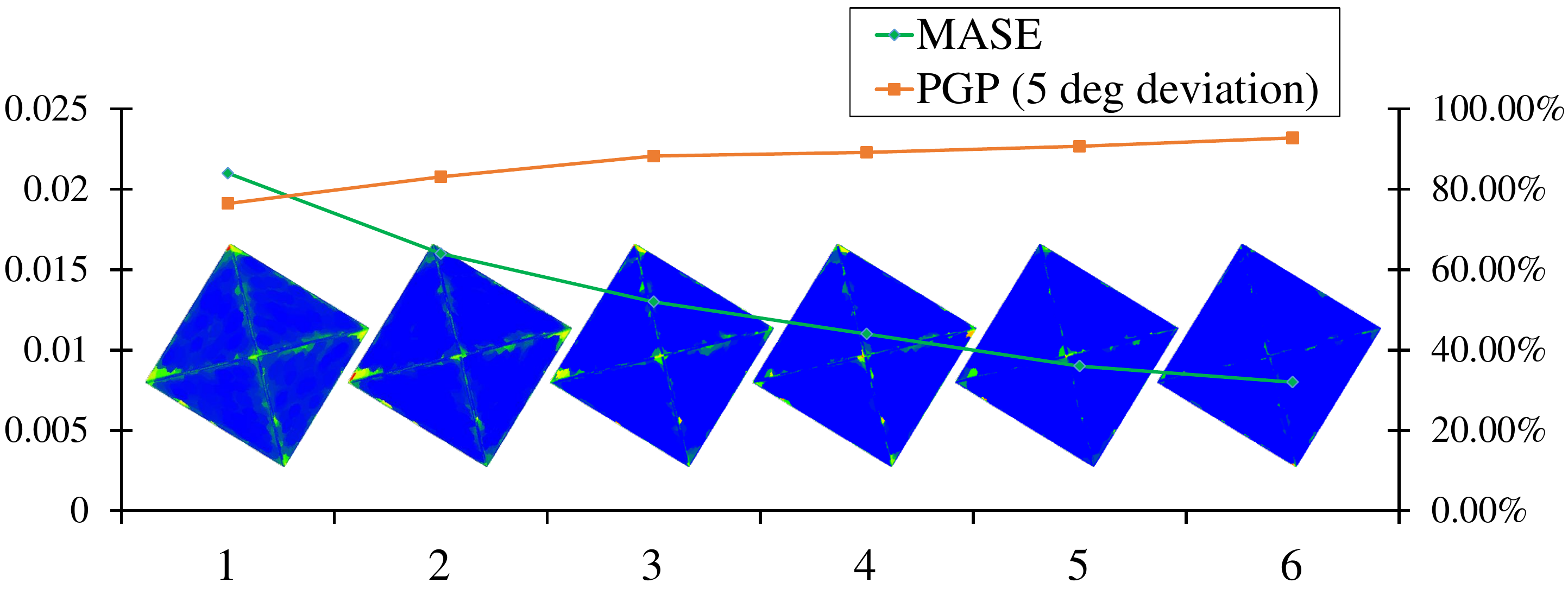}
  \caption{The normal estimation results in different iterations.
  MSAE: mean square angular error \cite{lu2018low}. PGP: the proportion of good points \cite{boulch2016deep}. Visual results describe the angle errors.
  \label{fig:normal_iteration}}
\end{figure}

\begin{figure}[h]
  \centering
  \includegraphics[width=\linewidth]{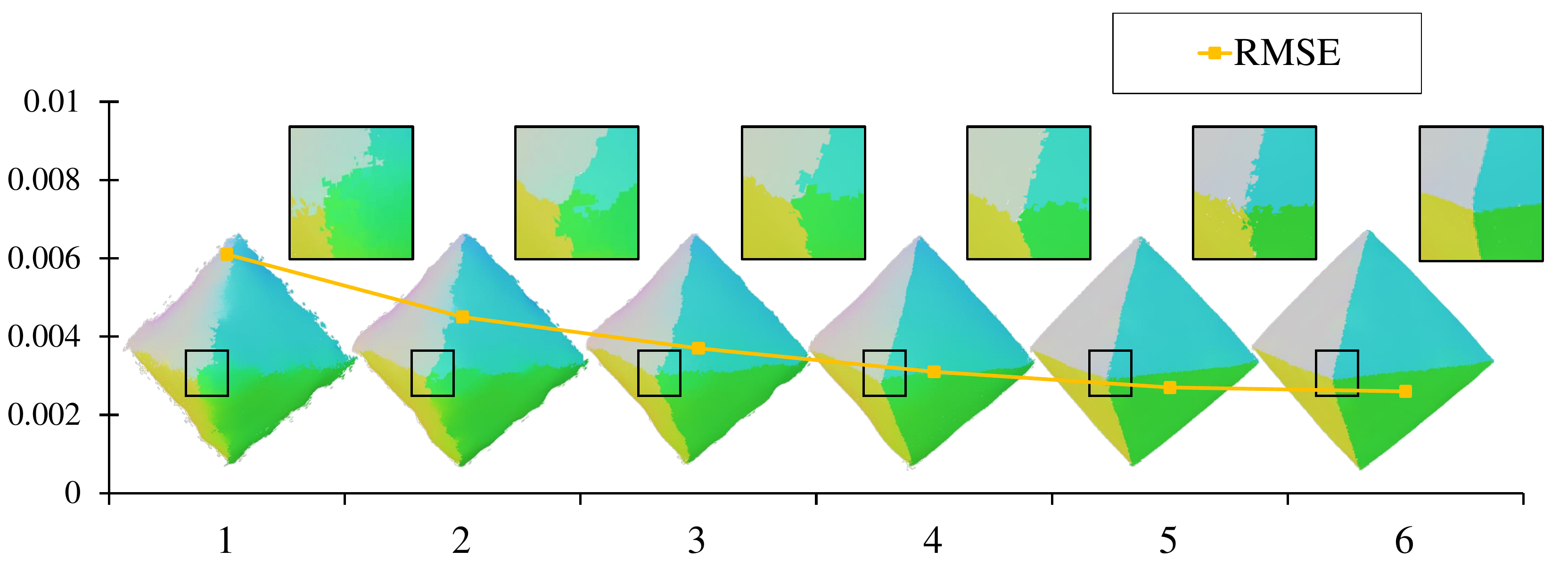}
  \caption{\ldn{The filtering results in different iterations. RMSE: root mean square error.}
  \label{fig:filtering_iteration}}
\end{figure}
To achieve better results for the testing phase, we alternately call the normal estimation and position update for a few iterations. The number of iterations $\kappa$ is empirically set to $6$, to mitigate the gap near sharp edges which has been pointed out in the original work \cite{lu2018low}. \ldn{As shown in Figure~\ref{fig:classification_iteration}, \ref{fig:normal_iteration} and \ref{fig:filtering_iteration}, as the number of iterations increases, the results of classification and normal estimation are getting better, which lead to better filtering results.
This is because the classification, normal estimation and position update work collaboratively. In other words, classification facilitates normal estimation and normal estimation facilitates position update, and position update further facilitates classification in next iteration. Therefore, the noise level of the input point cloud decreases with increasing iterations, and better results can be achieved.
}

\begin{figure}[h]
  \centering
  \includegraphics[width=\linewidth]{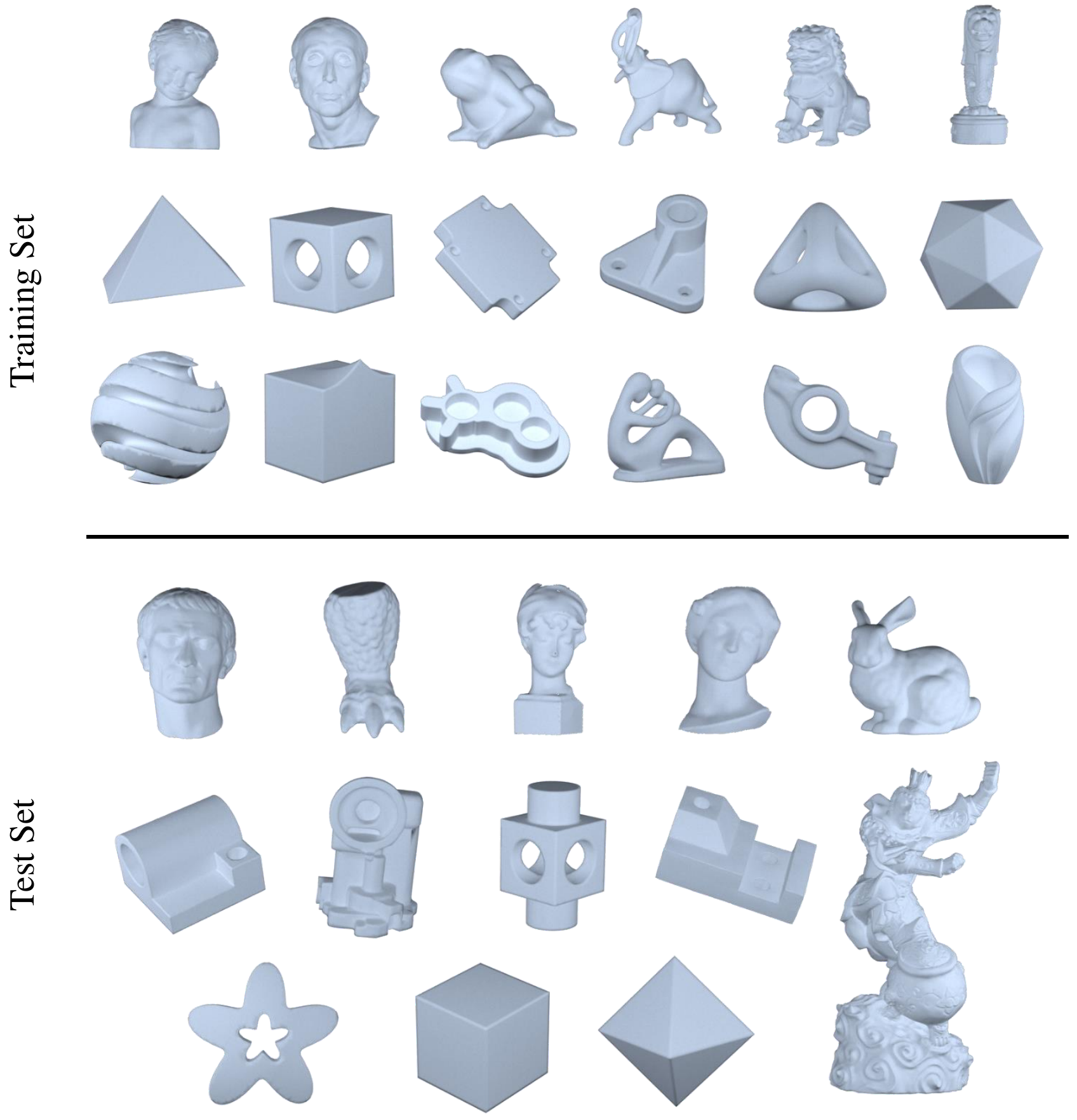}
  \caption{3D shapes used for training and testing.
  \label{fig:dataset}}
\end{figure}

\section{Results}
\label{sec:results}
We test our approach on a variety of raw and synthetic point clouds, and compare it with the state-of-the-art normal estimation approaches, including principal component analysis (PCA)~\cite{hoppe1992surface}, bilateral normal filtering (BF) \cite{huang2013edge}, Hough-based method (Hough) \cite{boulch2016deep} and Nesti-Net \cite{ben2019nesti}. Besides, we also compare our method with current point cloud filtering techniques, including EC-Net \cite{yu2018ec}, PointCleanNet \cite{rakotosaona2019pointcleannet}, CLOP \cite{preiner2014continuous}, RIMLS \cite{oztireli2009feature} and GPF \cite{lu2017gpf}.

For the sake of fair comparisons, the parameters of each method are tuned carefully, based on the suggested ranges or values. Similar to \cite{lu2017gpf}, we also upsample the denoised point cloud of the same input to a similar number of points for visualization purposes, and set the same parameter values for each method when reconstructing surfaces based on the upsampling results of the same model.

\begin{figure}[h]
  \centering
  \includegraphics[width=\linewidth]{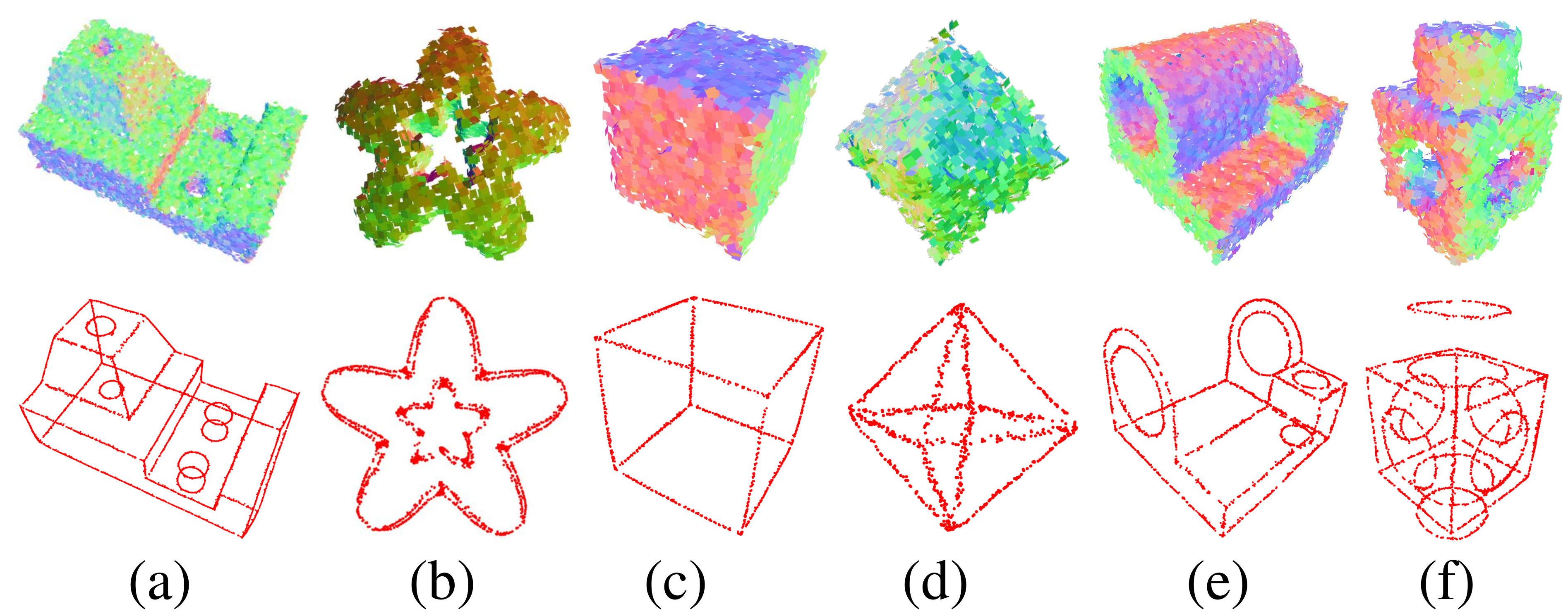}
  \caption{\ldn{Classification results on six CAD-like models. Classification accuracy ($\%$): (a) $97.2$, (b) $96.8$, (c) $98.9$, (d) $95.6$, (e) $97.0$ and (f) $93.7$.}
  \label{fig:classification_results}}
\end{figure}

\begin{figure}[b]
  \centering
  \includegraphics[width=\linewidth]{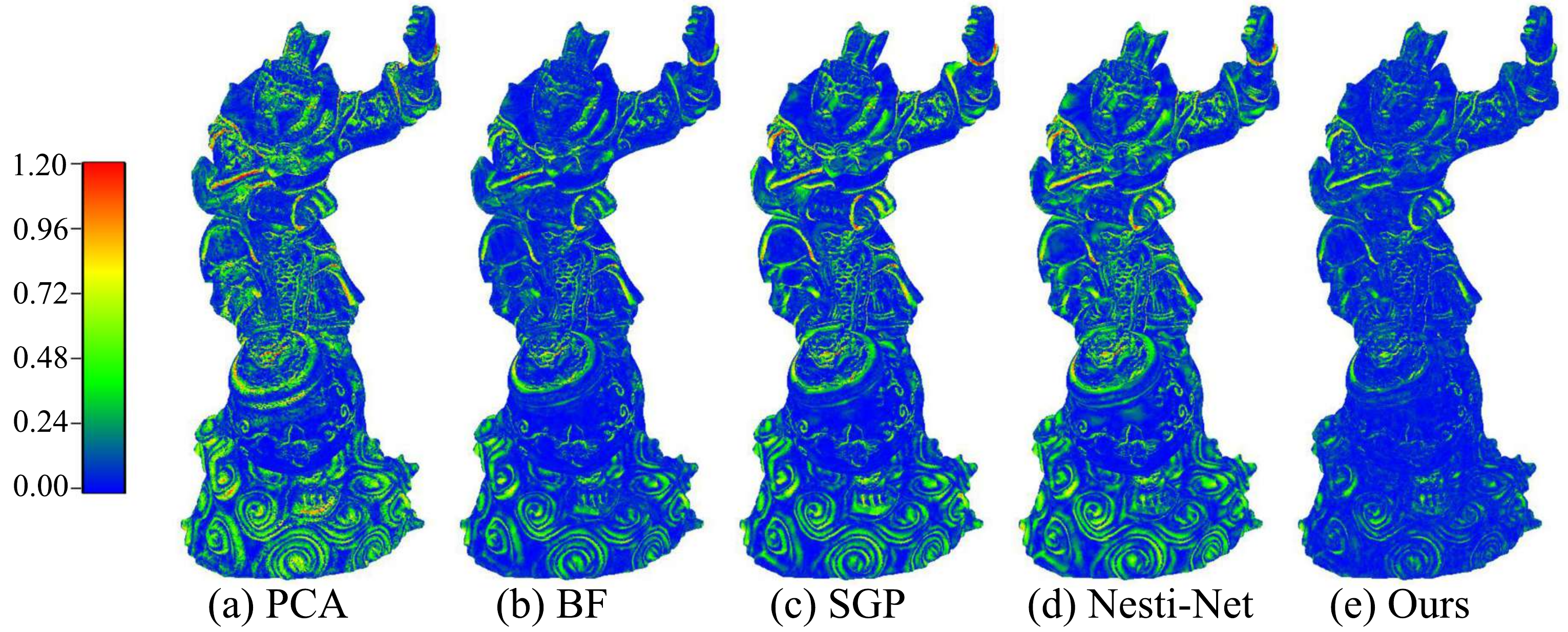}
  \caption{Normal estimation results on a noisy Monkey point cloud (0.5$\%$ noise).
  \label{fig:monkey}}
\end{figure}

\begin{figure}[h]
  \centering
    \subfigure{\includegraphics[width=0.45\textwidth]{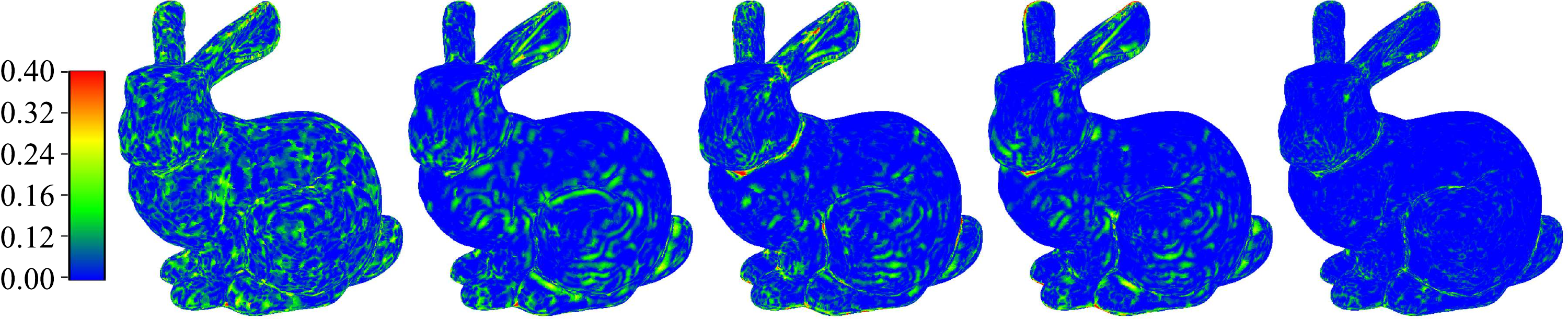}}\\
    \vspace{-4mm}
    \subfigure{\includegraphics[width=0.45\textwidth]{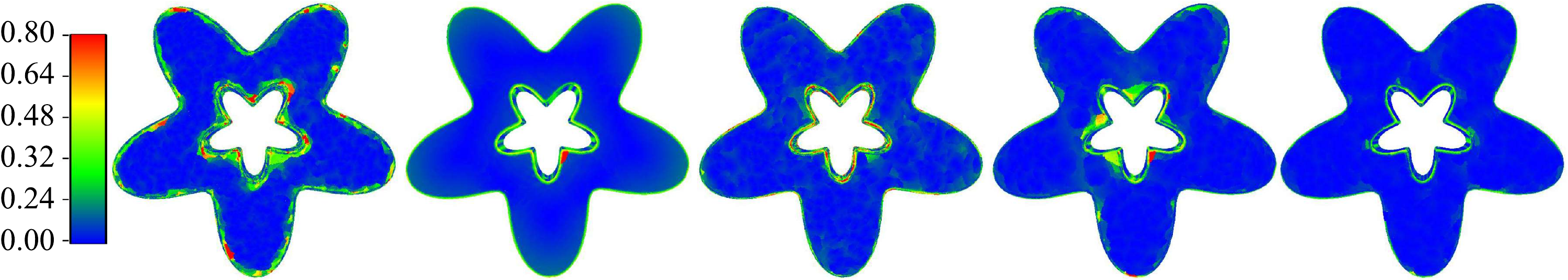}}\\
    \vspace{-4mm}
    \subfigure{\includegraphics[width=0.45\textwidth]{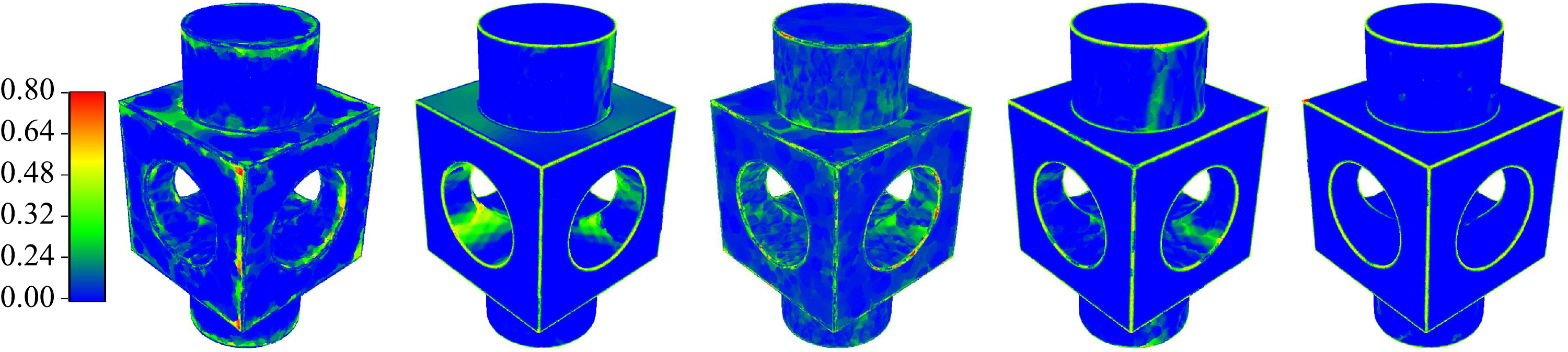}}\\
    \vspace{-4mm}
    \subfigure{\includegraphics[width=0.45\textwidth]{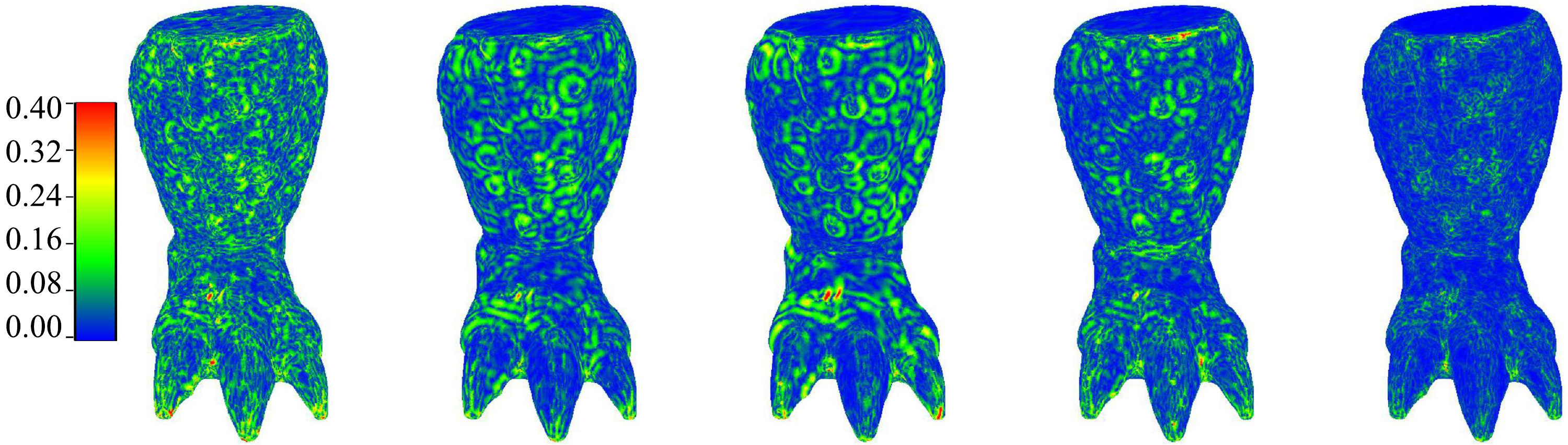}}\\
    \vspace{-4mm}
    \subfigure{\includegraphics[width=0.45\textwidth]{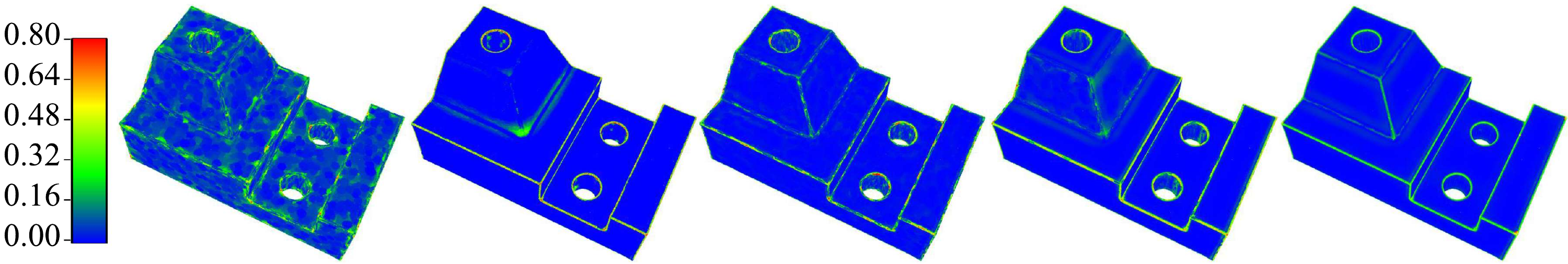}}\\
    \vspace{-4mm}
    \subfigure{\includegraphics[width=0.45\textwidth]{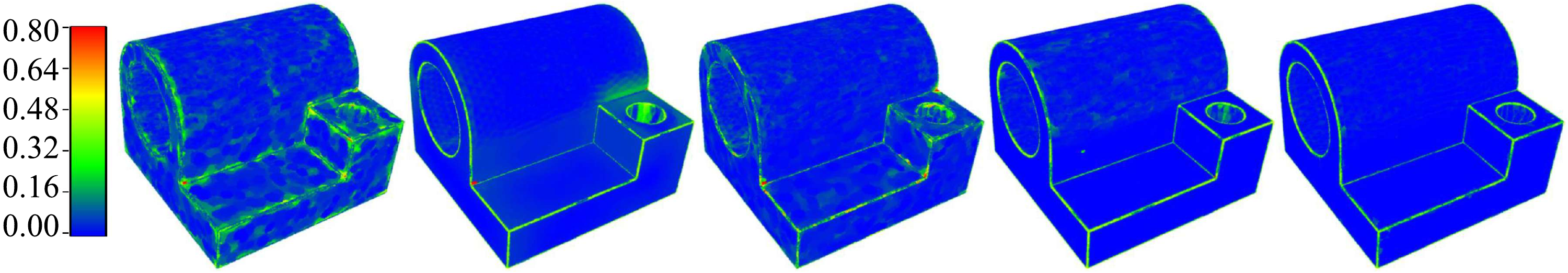}}\\
    \vspace{-4mm}
    \subfigure{\includegraphics[width=0.45\textwidth]{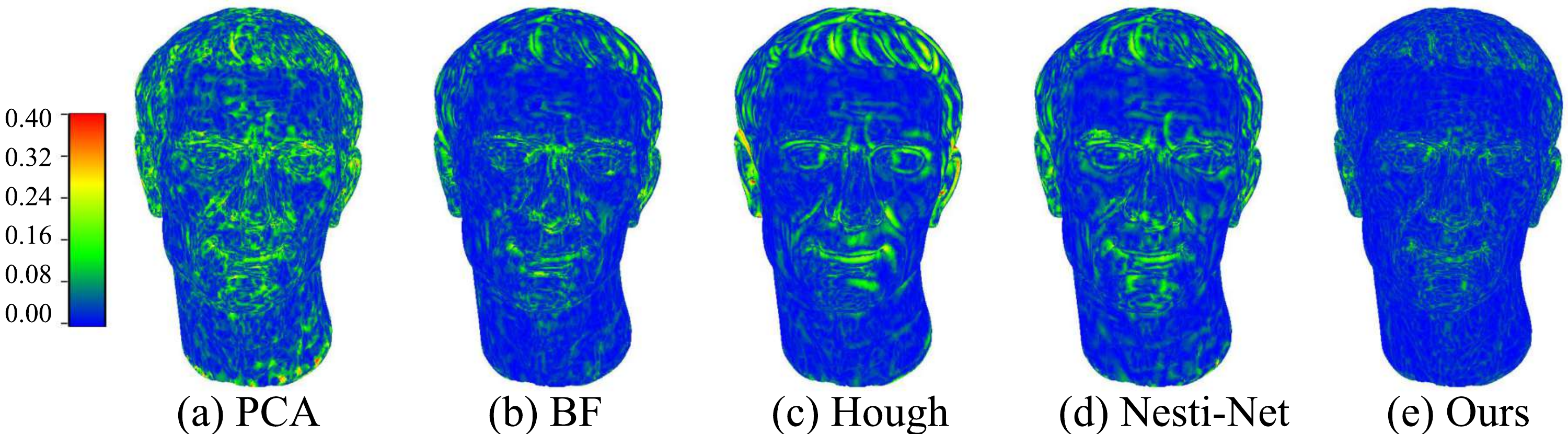}}
  \caption{Normal estimation results on seven models. From left to right: PCA \cite{hoppe1992surface}, BF \cite{huang2013edge}, Hough \cite{boulch2016deep}, Nesti-Net \cite{ben2019nesti} and Ours.
  }
    \label{fig:normal_quantiative}
\end{figure}

\begin{figure*}[ht]
  \centering
  \includegraphics[width=0.9\linewidth]{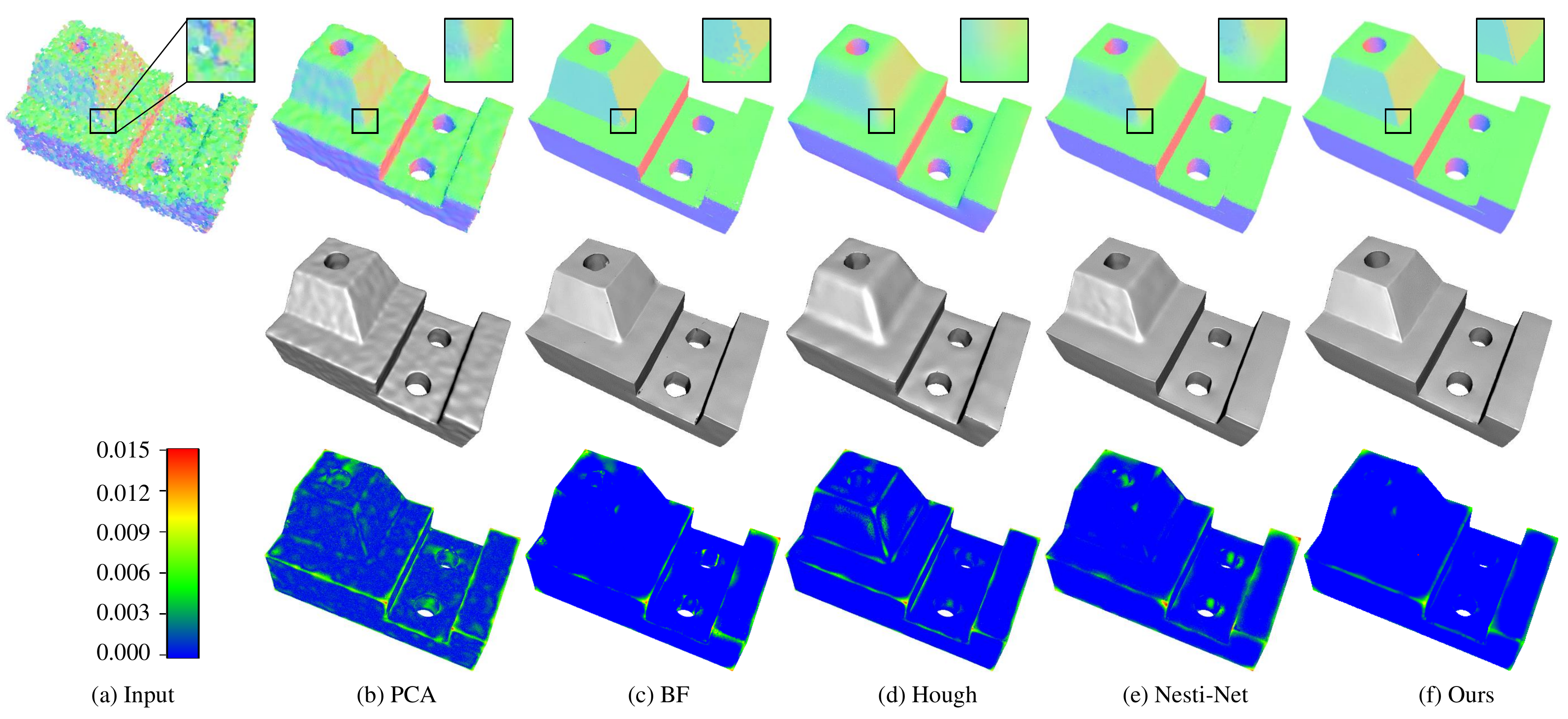}
  \caption{Filtering results on a noisy mechanical point cloud (1$\%$ noise). First row: filtering results with upsampling. Second row: corresponding surface reconstruction results. Third row: corresponding distance errors.
  The root mean square errors are ($\times 10^{-3}$): (b) $3.7$, (c) $2.4$, (d) $3.1$, (e) $2.7$ and (f) $1.9$.
  \label{fig:normal_update_model2}}
\end{figure*}

\begin{figure}[htbp]
  \centering
  \includegraphics[width=\linewidth]{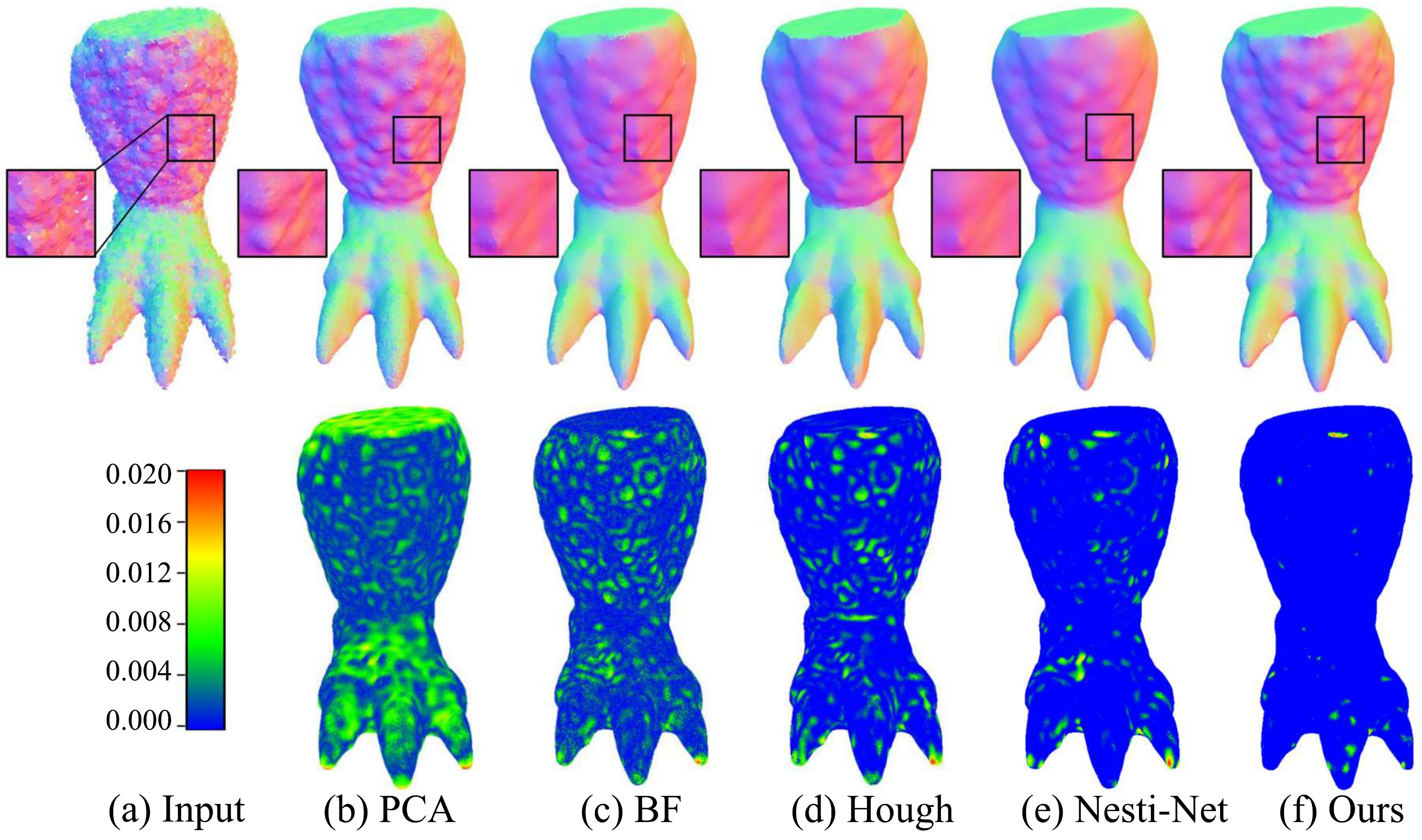}
  \caption{Filtering results on a noisy Leg point cloud (1$\%$ noise). First row: noisy input and filtering results with upsampling. Second row: distance errors.
  The root mean square errors are ($\times 10^{-3}$): (b) $6.9$, (c) $4.5$, (d) $2.5$, (e) $1.9$, (f) $1.3$.
  \label{fig:normal_update_13_nonmodeling}}
\end{figure}

\begin{figure}[hbp]
  \centering
  \includegraphics[width=\linewidth]{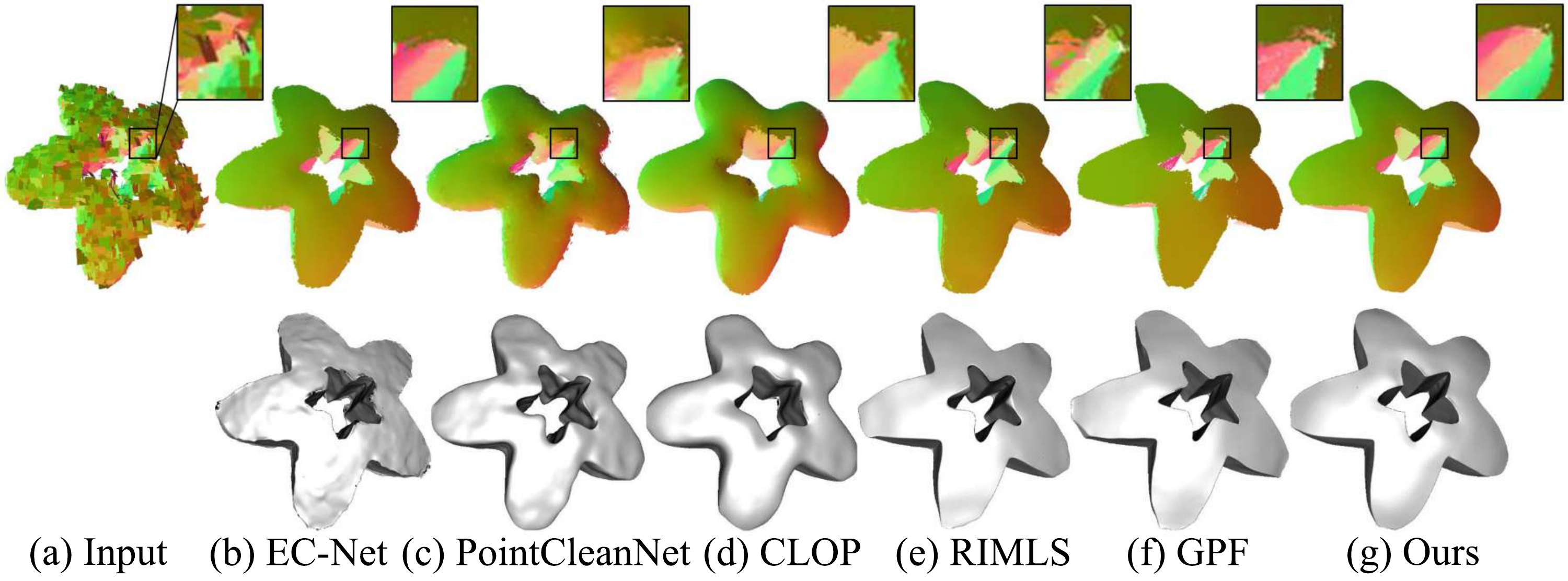}
  \caption{Filtering results on a noisy Trim-star point set (1$\%$ noise). First row: filtering results with upsampling. Second row: surface reconstruction results.
  \label{fig:tro_trim}}
\end{figure}

\begin{figure}[h]
  \centering
  \includegraphics[width=0.9\linewidth]{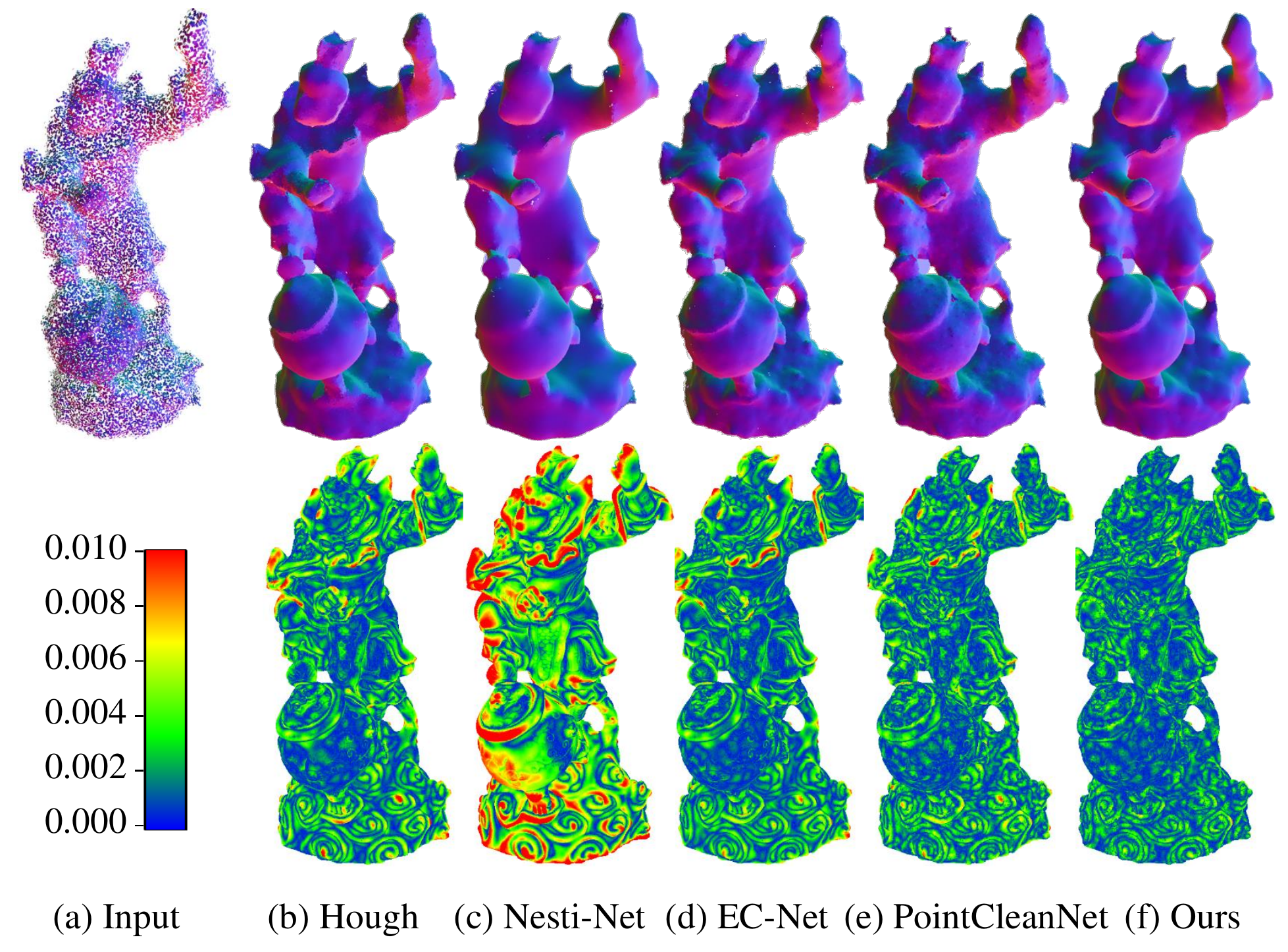}
  \caption{Filtering results on a noisy Monkey point cloud (0.5$\%$ noise). First row: filtering results with upsampling. Second row: distance errors.
  The root mean square errors are ($\times 10^{-3}$): (b) $2.0$, (c) $4.1$, (d) $1.9$, (e) $1.8$, (f) $1.5$.
  \label{fig:monkeyfiltering}}
\end{figure}

\subsection{Dataset and Parameter Setting}
\textbf{Dataset.}
Our collected dataset consists of $31$ models which have been split into $18$ training models and $13$ test models, shown in Figure \ref{fig:dataset}. The training set contains $11$ CAD-like models and $7$ smooth models which are all with synthetic noise.
For testing, we have $7$ CAD-like models and $6$ smooth models with different noise levels. We also test the real-world data captured by Kinect-v2 and Handy700 (included in the test set), to demonstrate the generalization capability of our method.
We generate the ground truth point clouds and corresponding normals from the triangle meshes via randomly sampling $10K-30K$ points. Each point cloud is corrupted with different levels of noise (Gaussian standard deviation $0.5\%$, $1\%$, $1.5\%$, $2\%$ of the bounding box diagonal length). We also augment the training data by applying a random rotation to each point cloud and swapping the axles of the eigenspace of each patch.
During training, the same number (2K in our experiments) of patches of each model are extracted randomly in an epoch,
to ensure that the training  can cover all types of models evenly. We train $100$ and $300$ epoches for the classification and normal estimation networks, respectively.

\textbf{Network implementation and parameters.}
We implemented the networks with Pytorch and trained them on a NVIDIA Geforce GTX 1080 GPU.
The classification and normal estimation networks are both trained with the Adam Optimizer, with a momentum of $0.9$ and starting with an initial learning rate of $0.001$ decreased by a factor of $0.9$ at every $100$ epoches.
\textit{All the involved parameters of our method are empirically set} (refer to Sec. \ref{sec:step1} - \ref{sec:step4}), to show the robustness of our approach.

\begin{table*}[htbp]
 \centering
\small
 \caption{Normal errors (mean square angular error, in radians) of the point clouds in Figure \ref{fig:normal_quantiative}.
 }
 \label{tab:normal_estimation}
 \begin{tabular}{ccccccccc}
  \hline
    \multirow{2}{*}{Models} & {Figure \ref{fig:normal_quantiative}} & {Figure \ref{fig:normal_quantiative}} & {Figure \ref{fig:normal_quantiative}} & {Figure \ref{fig:normal_quantiative}}  &{Figure \ref{fig:normal_quantiative}} & {Figure \ref{fig:normal_quantiative}} & {Figure \ref{fig:normal_quantiative}} \\
  & 1st row & 2nd row & 3rd row & 4th row & 5th row & 6th row & 7th row \\
 \hline

  \textbf{PCA}\cite{hoppe1992surface} & 0.051  & 0.086    & 0.097  & 0.017 & 0.091    & 0.040 & 0.052 \\
  \textbf{BF}\cite{huang2013edge}    & 0.006  & 0.024    & 0.022  & 0.006 & 0.009    & 0.007 & 0.018  \\
  \textbf{Hough}\cite{boulch2016deep}  & 0.009  & 0.046    & 0.043  & 0.007 & 0.013   & 0.033 & 0.026  \\
  \textbf{Nesti-Net}\cite{ben2019nesti} & 0.004  & 0.040    & 0.015  & 0.005 & 0.007    & 0.011 & 0.020 \\
  \textbf{Ours}    & \textbf{0.002}  & \textbf{0.007}    & \textbf{0.011}  & \textbf{0.002} & \textbf{0.005}   & \textbf{0.005} & \textbf{0.009}  \\
  \hline
 \end{tabular}
\end{table*}

\begin{figure*}[]
  \centering
    \subfigure{\includegraphics[width=\textwidth]{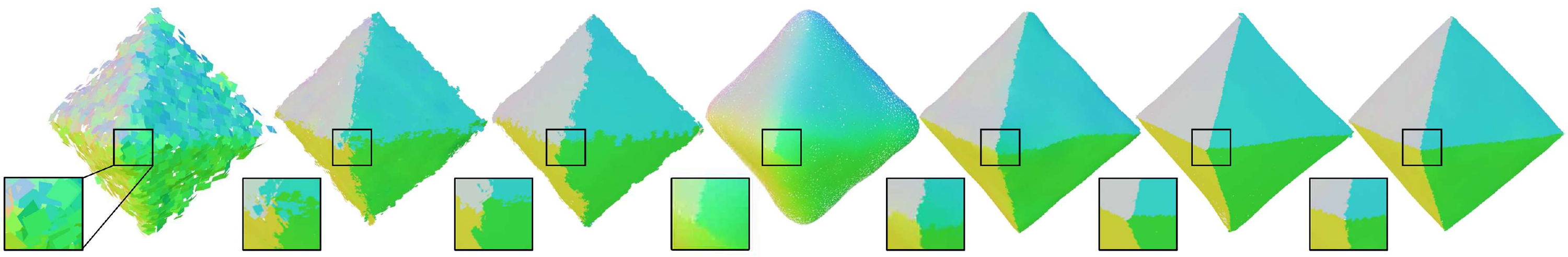}}\\
    \vspace{-4mm}
    \subfigure{\includegraphics[width=\textwidth]{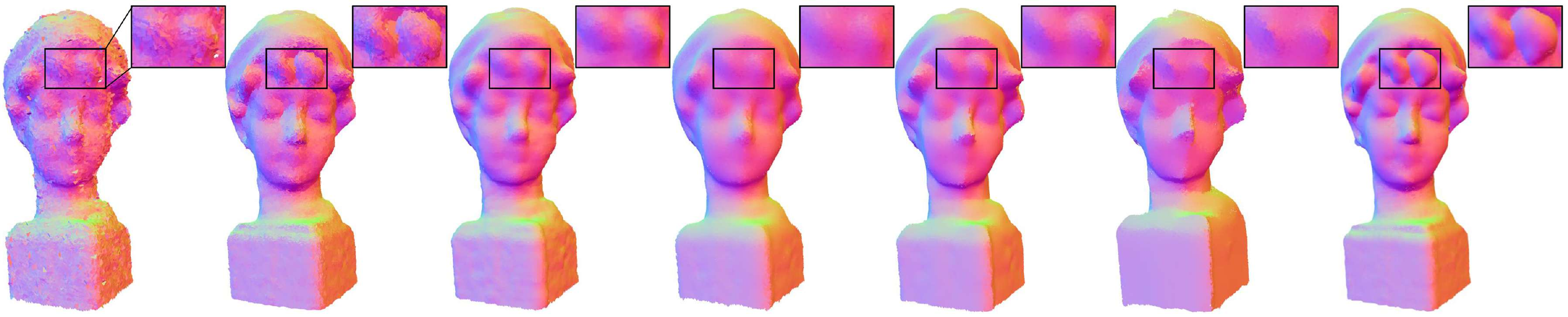}}\\
    \vspace{-4mm}
    \subfigure{\includegraphics[width=\textwidth]{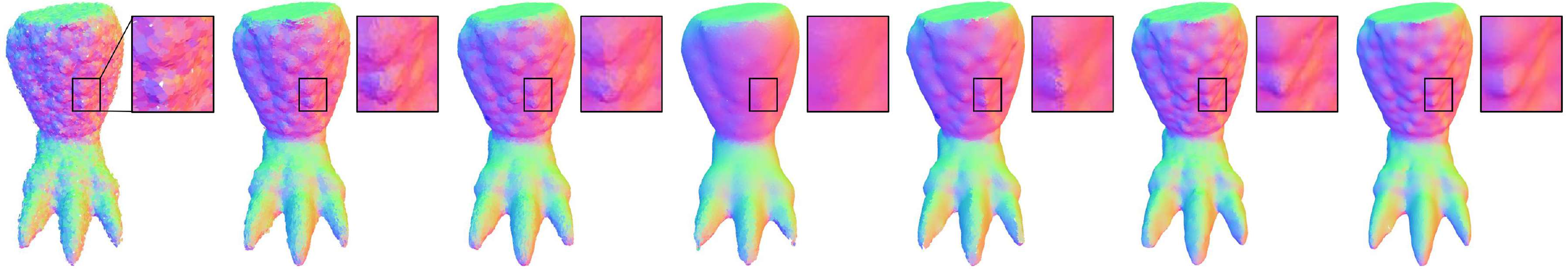}}\\
    \vspace{-4mm}
    \subfigure{\includegraphics[width=\textwidth]{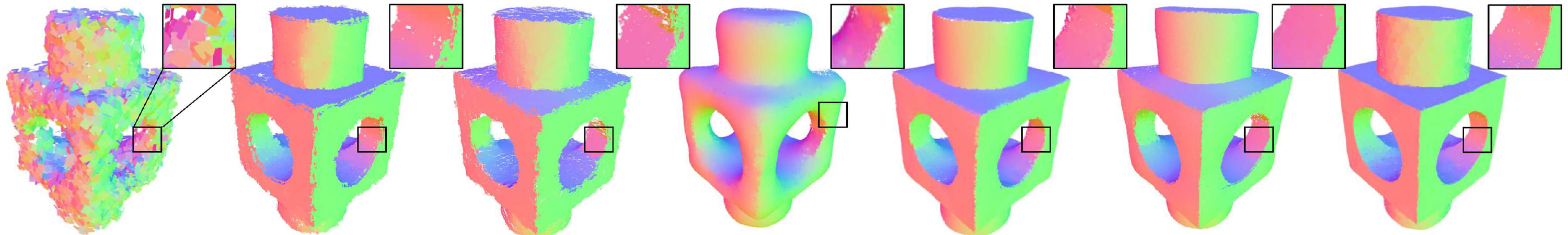}}\\
    \vspace{-4mm}
    \subfigure{\includegraphics[width=\textwidth]{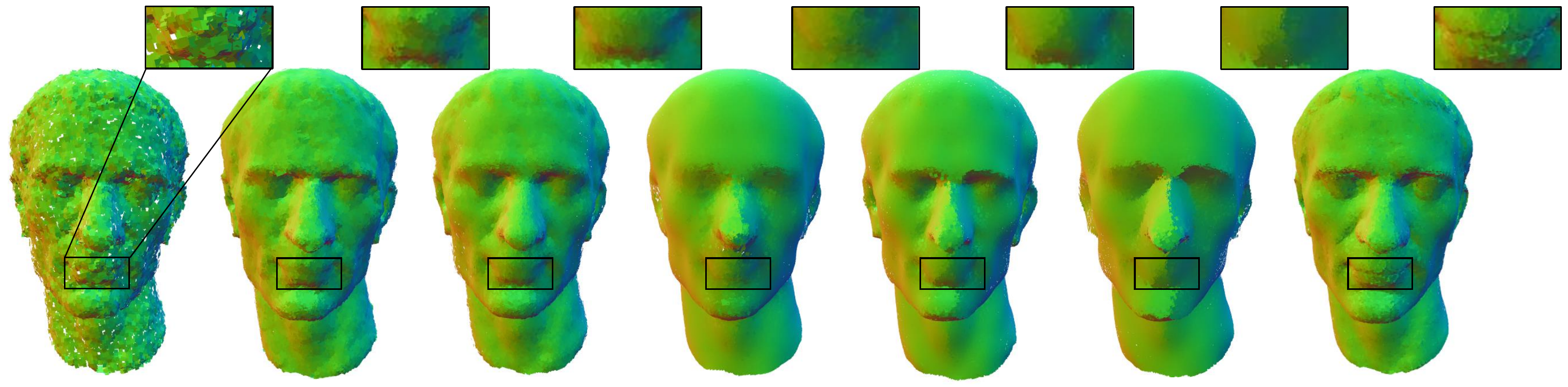}}\\
    \vspace{-4mm}
    \subfigure{\includegraphics[width=\textwidth]{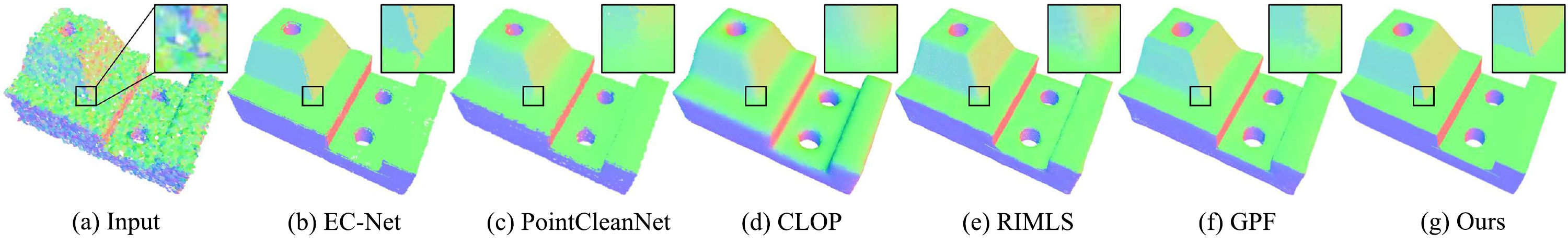}}\\
  \caption{Additional comparison of filtering results on seven point cloud models: Octahedron point set (1.5$\%$ noise), raw Girl point set captured by Kinect, Leg point set (1$\%$ noise), Block point set (1.5$\%$ noise),
  Julius point set (1$\%$ noise) and Joint point set (1$\%$ noise). From left to  right: noisy input and the filtering results of EC-Net \cite{yu2018ec}, PointCleanNet \cite{rakotosaona2019pointcleannet}, CLOP \cite{preiner2014continuous}, RIMLS \cite{oztireli2009feature}, GPF \cite{lu2017gpf} and Ours.
}
    \label{fig:tro_sub}
\end{figure*}

\begin{figure*}[ht]
  \centering
  \includegraphics[width=0.9\linewidth]{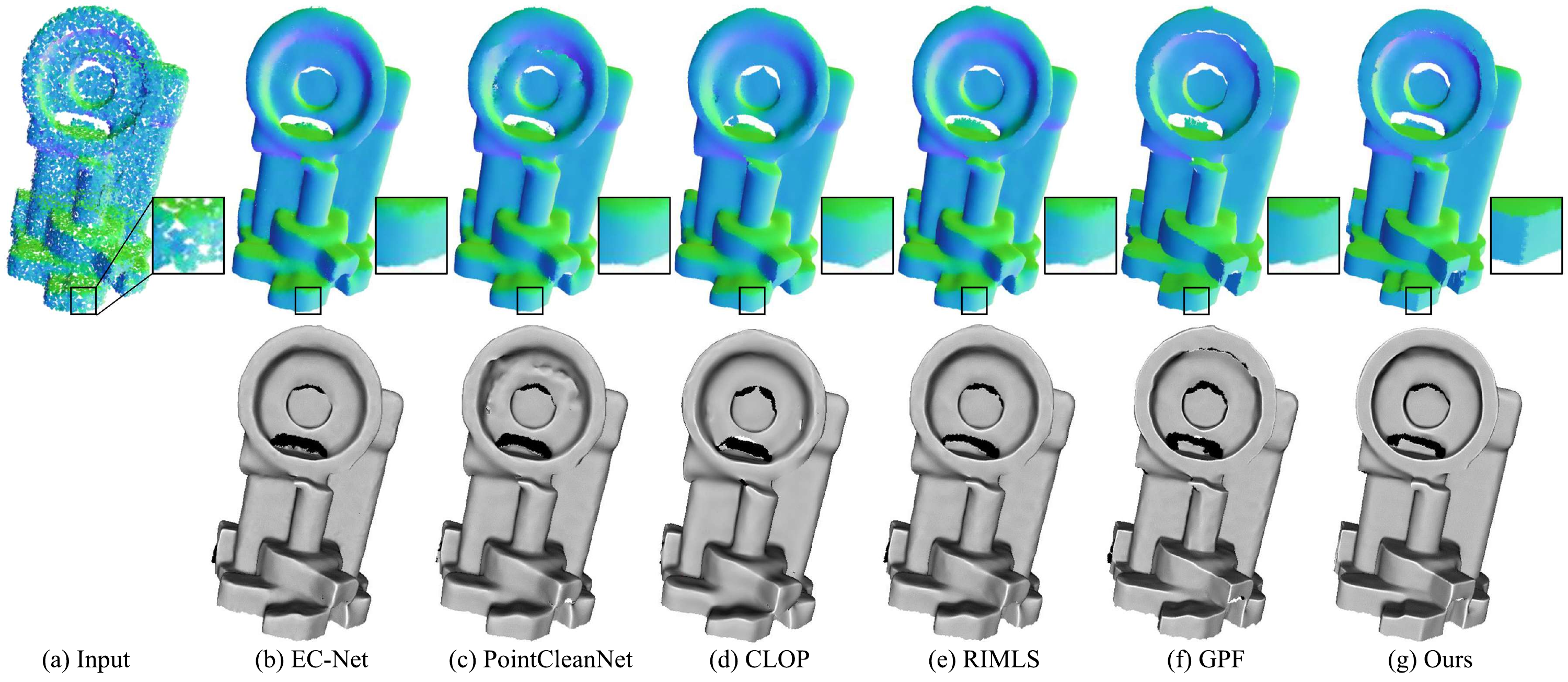}
  \caption{Filtering results on a noisy point cloud captured by Handy700. First row: filtering results with upsampling. Second row: surface reconstruction results.
  \label{fig:tro_real_scan_1}}
\end{figure*}

\begin{figure}[h]
  \centering
  \includegraphics[width=0.9\linewidth]{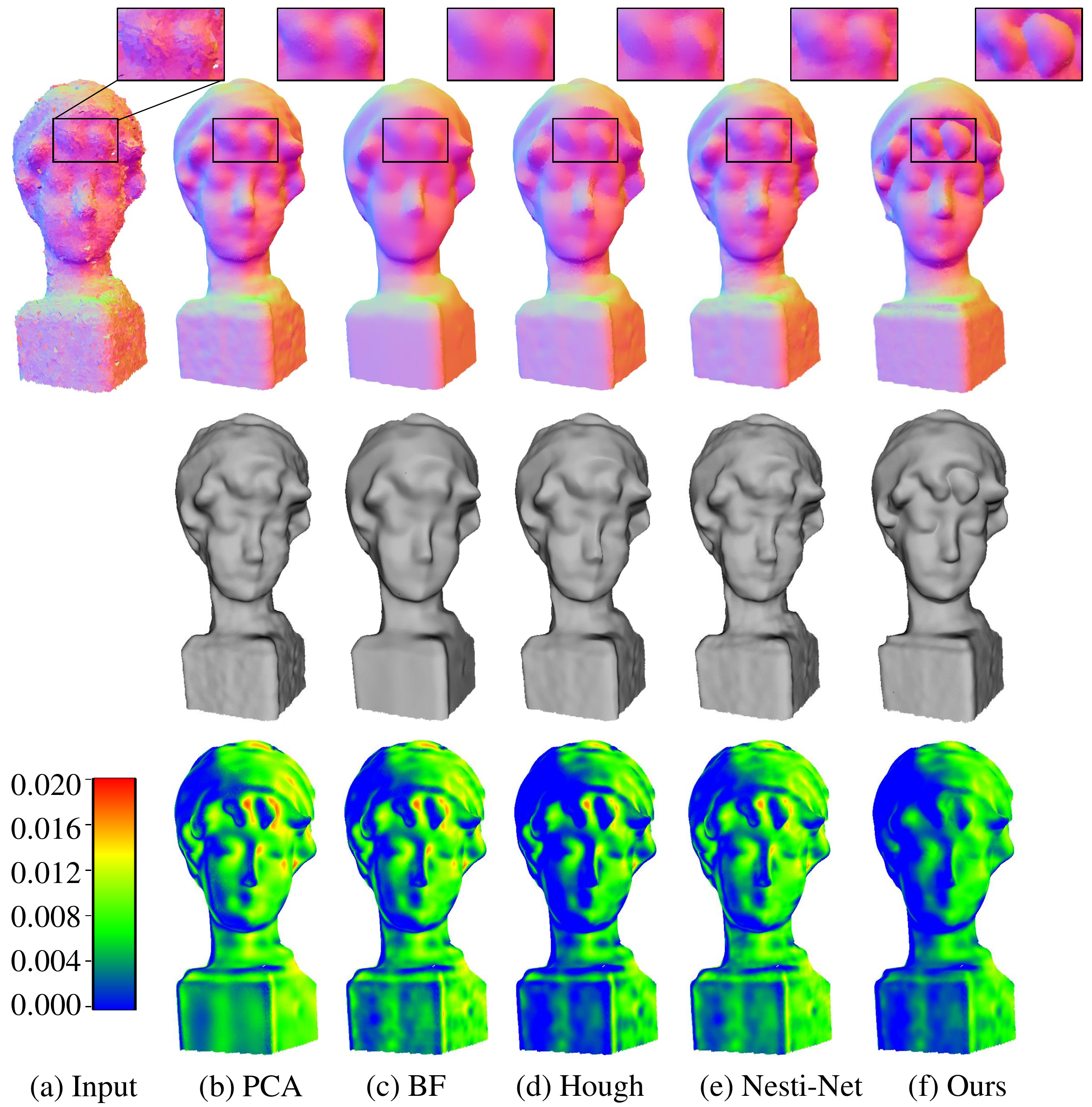}
  \caption{Filtering results on a noisy Girl point set obtained by Kinect. First row: filtering results with upsampling. Second row: surface reconstruction results. Third row: distance errors.
  The root mean square errors are ($\times 10^{-3}$): (b) $9.3$, (c) $8.4$, (d) $6.5$, (e) $8.2$, (f) $5.6$.
  \label{fig:normal_update_girl}}
\end{figure}

\subsection{Normal Estimation}
In general, PCA \cite{hoppe1992surface} estimates the normal of a point by fitting a local plane to its neighbors. Thus, the estimated normals tend to smooth sharp features. Bilateral normal smoothing \cite{huang2013edge} is suitable for CAD-like models, but prone to over-sharpen some regions to reach the overall smoothing effect. The Hough-based method \cite{boulch2016deep} has challenges in predicting normals of sharp feature points, since it does not consider the distinctions of feature and non-feature points. Given fisher vectors as input, the Nesti-Net~\cite{ben2019nesti} relies on a data driven approach for selecting the optimal scale around each point and encourages sub-network specialization. It works relatively poorly in a single scale. Compared with these methods, our normal estimation approach treats feature and non-feature points separately, and achieves the best results (i.e., lowest errors), as shown in Table~\ref{tab:normal_estimation}.

\subsection{Point Cloud Filtering}
Besides the above normal estimation, we also demonstrate the proposed approach in the point cloud filtering application.

\begin{figure}[h]
  \centering
  \includegraphics[width=\linewidth]{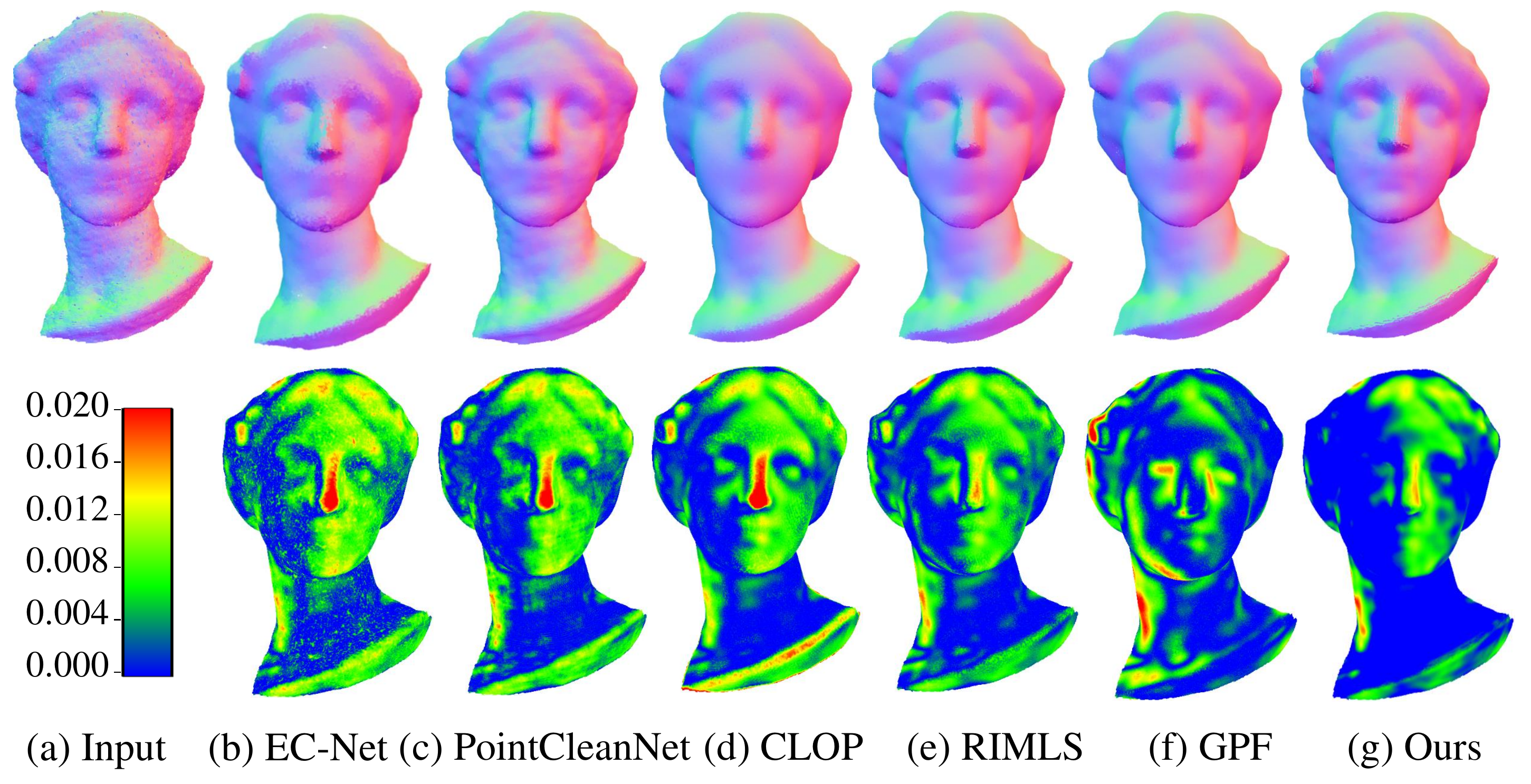}
  \caption{Filtering results on a noisy point cloud obtained by Kinect (Big Girl). First row: filtering results with upsampling. Second row: distance errors. 
  The root mean square errors are ($\times 10^{-3}$): (b) $7.3$, (c) $7.0$, (d) $7.1$, (e) $6.5$, (f) $5.7$, (g) $4.4$. 
  \label{fig:tro_big_girl}}
\end{figure}

\textbf{Synthetic point clouds.}
We apply the position update algorithm \cite{lu2018low} to match the estimated normals output by the above normal estimation methods. \textit{Generally, a better normal estimation result will lead to a more accurate denoised model}. Figures  \ref{fig:normal_update_model2}, \ref{fig:normal_update_13_nonmodeling} and \ref{fig:monkeyfiltering} show that our normal estimation method can enable better point cloud filtering results than state-of-the-art normal estimation approaches.
Apart from the filtering comparisons with the normal estimation methods, we also compare our method with the state-of-the-art denoising methods, including EC-Net \cite{yu2018ec}, PointCleanNet \cite{rakotosaona2019pointcleannet}, CLOP \cite{preiner2014continuous}, RIMLS \cite{oztireli2009feature} and GPF \cite{lu2017gpf}.

In Figure \ref{fig:tro_trim}, RIMLS tends to smooth out some small-scale features.
GPF generates an over-sharpened result near the cylindrical features, and leads to an uneven distribution in the denoised model due to the edge absorption. Since the EC-Net and PointCleanNet are independent of normals, we use the bilateral normal smoothing \cite{huang2013edge} before their reconstruction, by setting the same parameters. This actually enhances their reconstruction results. Figure \ref{fig:monkeyfiltering} shows the comparisons of the deep learning methods on a complicated Monkey point cloud, and Figure \ref{fig:tro_sub} shows the results of different point cloud filtering techniques over six different models, which also confirms the outstanding performance of our method. 
Since we classify points to features and non-feature points and handle them separately, our method achieves the best results in the preservation of both straight and curved sharp edges among all methods.

\begin{figure}[htbp]
  \centering
  \includegraphics[width=0.8 \linewidth]{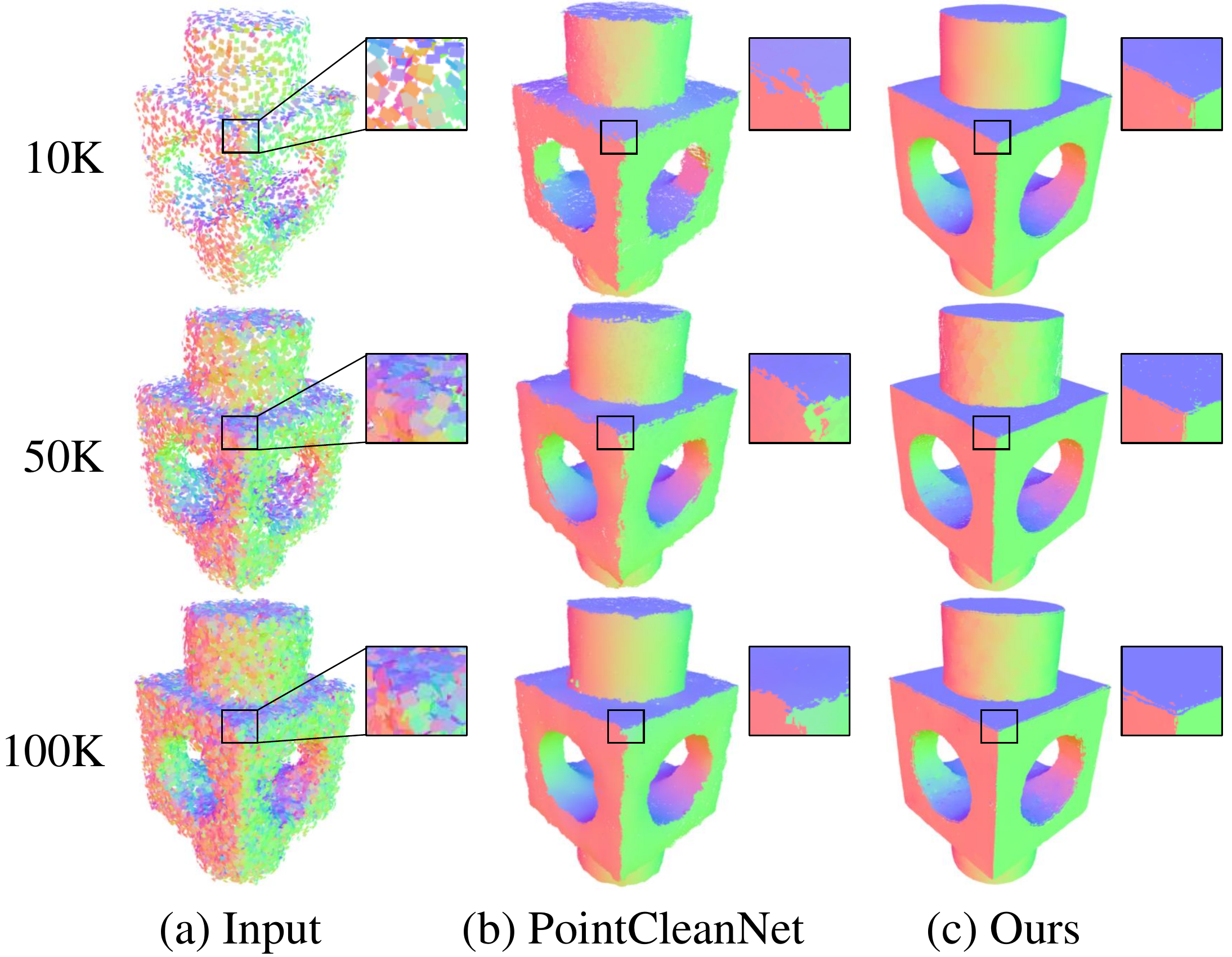}
  \caption{Filtering results of of our method and PointCleanNet \cite{rakotosaona2019pointcleannet} over the Block point clouds with different densities.
  From the first row to the third row: 10K, 50K and 100K points, respectively.
  \label{fig:robustness1}}
\end{figure}

\begin{figure}[h]
  \centering
  \includegraphics[width=0.9\linewidth]{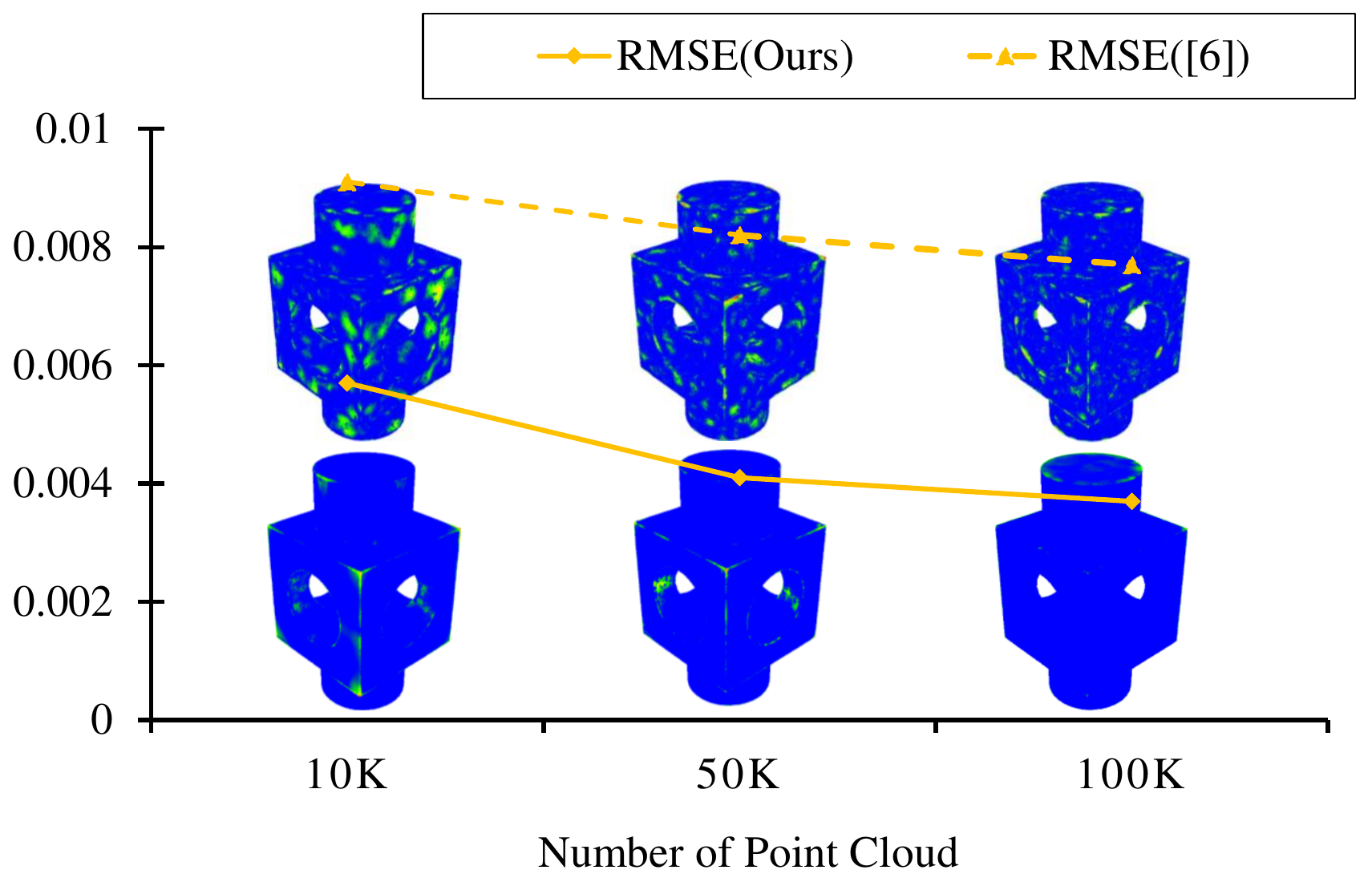}
  \caption{Distance errors of our method and PointCleanNet \cite{rakotosaona2019pointcleannet} over the Block point clouds with different densities.
  \label{fig:robustness2}}
\end{figure}

\textbf{Raw point scans.}
Besides the synthetic models, we further contrast our approach with these methods on the scanned data with raw noise. Figure \ref{fig:tro_real_scan_1}, \ref{fig:normal_update_girl} and \ref{fig:tro_big_girl} show the filtering results by updating positions based on normal estimation and the state-of-the-art point set filtering methods. We can see that the results by our approach have higher fidelity to the ground truth than other methods, with regard to preserving features.

\begin{table*}[htbp]
 \centering
\small
 \caption{Chamfer measure ($\times 10^{-5}$) of the point clouds in Figure \ref{fig:tro_sub}.
 }
 \label{tab:chamfer_distance}
 \begin{tabular}{cccccccc}
  \hline
    \multirow{2}{*}{Models} & {Figure \ref{fig:tro_sub}} & {Figure \ref{fig:tro_sub}} & {Figure \ref{fig:tro_sub}} & {Figure \ref{fig:tro_sub}}  &{Figure \ref{fig:tro_sub}} & {Figure \ref{fig:tro_sub}}\\
  & 1st row & 2nd row & 3rd row & 4th row & 5th row & 6th row \\
 \hline

  \textbf{EC-Net}\cite{yu2018ec} & 4.59  & 10.72    & 1.25  & 12.41 & 9.73    & 1.68\\
  \textbf{PointCleanNet}\cite{rakotosaona2019pointcleannet}    & 5.38  & 12.20    & 1.54  & 14.73 & 8.41    & 1.84 \\
  \textbf{CLOP}\cite{preiner2014continuous}  & 5.75  & 14.15    & 2.37  & 16.92  & 13.11   & 2.26  \\
  \textbf{RIMLS}\cite{oztireli2009feature} & 2.16  & 13.30    & 1.47  & 5.94 & 11.37    & 1.44 \\
  \textbf{GPF} \cite{lu2017gpf}   & 1.73  & 16.88    & 1.10  & 4.73 & 11.82    & 1.24  \\
  \textbf{Ours}    & \textbf{1.14}  & \textbf{6.33}    & \textbf{0.59}  & \textbf{3.37}& \textbf{5.81} & \textbf{0.77} \\
  \hline
 \end{tabular}
\end{table*}

\subsection{Quantitative Results}
To quantitatively evaluate the normal estimation and denoising quality, we utilize several published metrics to measure the results output by the involved methods, against the corresponding ground truth. \ldn{We measure the accuracy for classification, which is described as the proportion of the true feature points in the predicted feature point set. Figure \ref{fig:classification_loss_weight}, \ref{fig:feature_move}, \ref{fig:classification_iteration} and \ref{fig:classification_results} show the classification results both visually and quantitatively. It demonstrates that classification is useful and necessary to the normal estimation, and a higher accuracy would typically lead to a better normal estimation result.
}
For normal estimation, we evaluate the normal errors by adopting the MSAE (mean square angular error) \cite{lu2018low},
as listed in Table~\ref{tab:normal_estimation}. 
Figure \ref{fig:monkey} visualizes the normal errors for some test models, and our normal estimation result is better than those by the state-of-the-art techniques. This is also confirmed by the comprehensive normal estimation results shown in Figure \ref{fig:normal_quantiative}.
For point cloud filtering, we calculate the mean distances between points of the ground truth and their closet points of the upsampled point cloud \cite{lu2017gpf}, and compute the RMSE metric of all mean distances in a point cloud to measure the geometric fidelity.
\ldnc{In addition, we also calculate the \textit{Chamfer distance} \cite{rakotosaona2019pointcleannet} in Table~\ref{tab:chamfer_distance}.}
Figures \ref{fig:normal_update_model2}, \ref{fig:normal_update_13_nonmodeling}, \ref{fig:monkeyfiltering}, \ref{fig:normal_update_girl} and \ref{fig:tro_big_girl} include a visualization of the distance errors, respectively. The distance errors show that our normal estimation method can produce obviously better filtering results than those by state-of-the-art normal estimation and point cloud filtering methods.

\subsection{Robustness to Density Variation}
Both our train set and test set contains noisy data with different densities. We note that some methods like PointCleanNet \cite{rakotosaona2019pointcleannet} are designed to process dense point clouds, which would usually produce less pleasant results on the sparse-density data. In comparison, our method employs an adaptive radius and Gaussian-weight interpolation for the height maps generation,
and the train set comprises of point clouds with different densities. As such, it is able to flexibly deal with different densities.
As shown in Figures \ref{fig:robustness1} and \ref{fig:robustness2}, taking the block as an example, we provide the visual and quantitative comparisons on varied densities for PointCleanNet \cite{rakotosaona2019pointcleannet} and our method. We can see that both methods can obtain better results for the dense point clouds (50K/100K points), compared to the less dense data (10K points). Our method generates consistently better results than PointCleanNet \cite{rakotosaona2019pointcleannet} for all densities, which manifests that our method is more robust to density variation.

\subsection{Limitations}
\begin{figure}[htbp]
  \centering
  \includegraphics[width=0.8\linewidth]{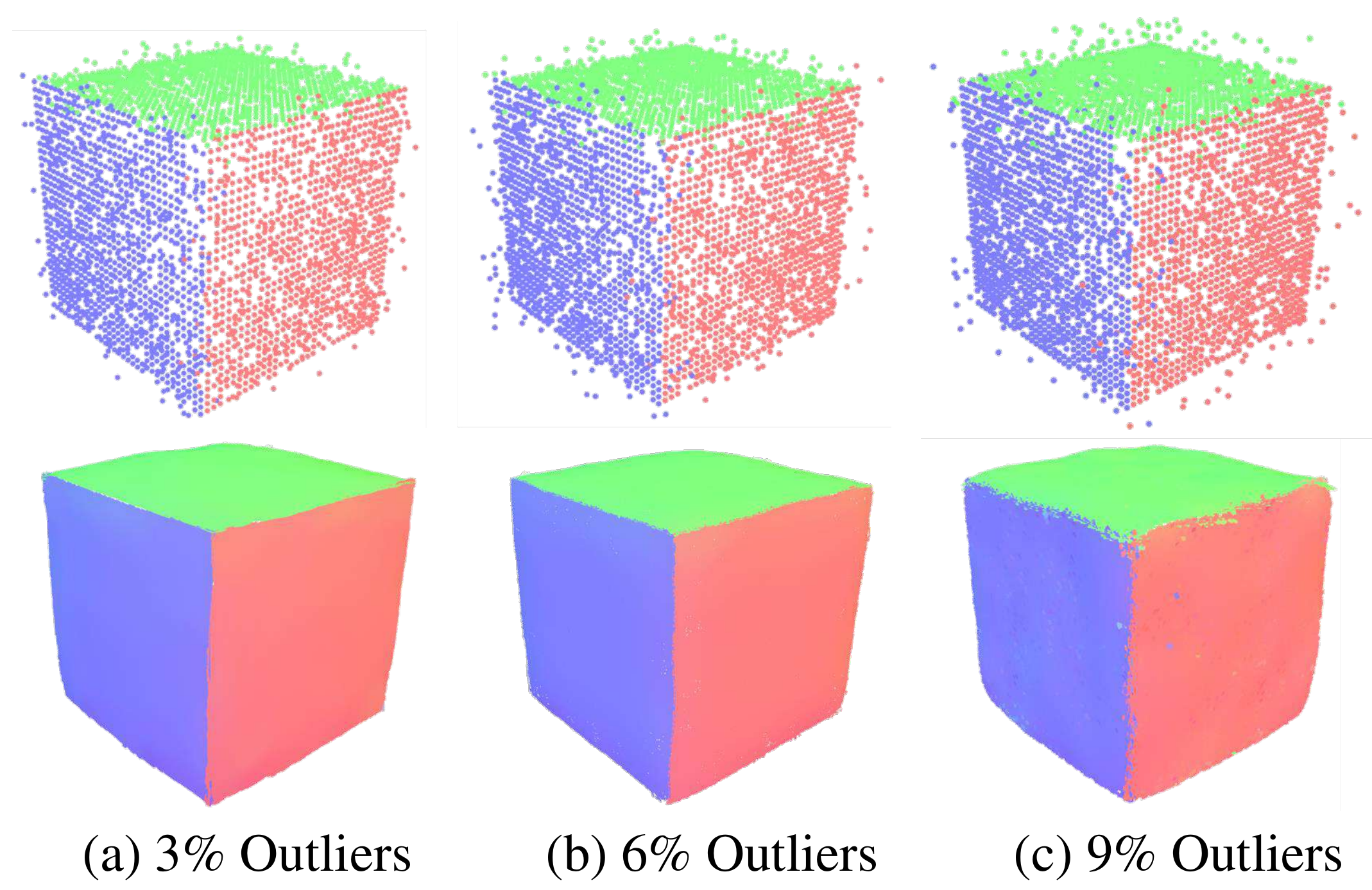}
  \caption{Filtering results of of our method over the Cube point clouds with different levels of outliers. First row: the inputs with $3\%$, $6\%$ and $9\%$ outliers. Second row: the filtering results with upsampling.
  \label{fig:outlier}}
\end{figure}

Despite the above advantages of our method, it still has a few limitations. 
Similar to existing works \cite{lu2017gpf,lu2018low}, our approach has difficulty in handling exceptionally severe noise. This is because such noise would significantly destroy the underlying surfaces, and lead to many ambiguities for feature and non-feature points. Second, our method is not particularly designed to deal with significant outliers. Figure \ref{fig:outlier} shows that  our approach is capable of handling the $3\%$ and $6\%$ outliers, and generally produce less desired results in the presence of heavier outliers ($9\%$). Our method and PointCleanNet \cite{rakotosaona2019pointcleannet} might complement each other, in terms of features preservation and outliers removal. In the future, we would like to investigate and solve the above issues by introducing innovative techniques.
\ldnc{Points and height maps are two different representations which satisfy the PointNet form and 2D CNNs, respectively. We would also like to exploit the two representations further in the future. 
}

\section{Conclusion}
\label{sec:conclusion}
In this paper, we have proposed a novel method for feature-preserving normal estimation and further point cloud filtering. Taking a noisy point cloud as input, our approach can automatically estimate its normals. Point positions are then updated to match the estimated normals. To achieve decent filtering results, we iterate the normal estimation and position update for a few times. Extensive experiments prove the effectiveness of our method, and demonstrate that it is both visually and quantitatively better than state-of-the-art normal estimation techniques and point cloud filtering methods.


\section*{References}
\bibliography{mybibfile}

\end{document}